\documentclass[english,journal,draftclsnofoot,onecolumn]{IEEEtran}
\usepackage[T1]{fontenc}
\usepackage[latin9]{inputenc}
\usepackage{float}
\usepackage{amsmath}
\usepackage{amssymb}
\usepackage{stackrel}
\usepackage{graphicx}

\makeatletter

\newcommand{\lyxmathsym}[1]{\ifmmode\begingroup\def\b@ld{bold}
  \text{\ifx\math@version\b@ld\bfseries\fi#1}\endgroup\else#1\fi}

\floatstyle{ruled}
\newfloat{algorithm}{tbp}{loa}
\providecommand{\algorithmname}{Algorithm}
\floatname{algorithm}{\protect\algorithmname}

\@ifundefined{date}{}{\date{}}
\@ifundefined{showcaptionsetup}{}{%
 \PassOptionsToPackage{caption=false}{subfig}}
\usepackage{subfig}
\makeatother

\usepackage{babel}
\begin{document}
\title{A Fast Iterative Algorithm to design phase only sequences by minimizing
the ISL metric }
\author{Surya Prakash Sankuru, Prabhu Babu}
\maketitle
\begin{abstract}
Unimodular/Phase only sequence having impulse like aperiodic auto-correlation
function plays a central role in the applications of RADAR, SONAR,
Cryptography, and Wireless (CDMA) Communication Systems. In this paper,
we propose a fast iterative algorithm to design phase only sequences
of arbitrary lengths by minimizing the Integrated Side-lobe Level
(ISL) metric, which is very closely related to the auto-correlation
property of a sequence. The ISL minimization problem is solved iteratively
by using the Majorization-Minimization (MM) technique, which ensures
a monotonic convergence to the stationary minimum point. To highlight
the performance of a proposed algorithm, we conduct the numerical
experiments for different sequence lengths using different initializations
and also compare them with the existing algorithms. Numerical simulations
show that irrespective of the sequence length and initialization,
the proposed algorithm is performing better than the state-of-the-art
algorithms in terms of speed of convergence. We also show a computationally
efficient way to implement our proposed algorithm by using the FFT
and IFFT operations.

Index Terms-- Majorization-Minimization, Integrated Side-lobe Level,
Peak Side-lobe Level, Phase only sequence, aperiodic auto-correlation,
Cryptography, Communication Systems, RADAR, SONAR, MIMO RADAR, CDMA.
\end{abstract}

\section*{\centerline{I.INTRODUCTION}}

The target detection capability of an active sensing system will solely
depend on the accuracy of the estimated underlying parameters. So,
to increase the detection performance in applications like active
sensing systems (SONAR, RADAR) \cite{1_DSP_SONAR}, \cite{2_RadarHandbook},
\cite{3_Radarsignal_Levanon}, \cite{Opt_Activesen}, \cite{Activesonar},
Cryptography \cite{4_signal_App}, CDMA communication systems \cite{7_probingwaveformrxsyn_robertHeStoica},
\cite{Opt_Rad_comm}, \cite{CDMA}, and MIMO RADAR {[}10-17{]},\nocite{diversity_iet_text}\nocite{khalili_cloud}\nocite{MIMO_lp}\nocite{MRBELL}\nocite{MULRADAR}\nocite{Yong_eusipco}
\nocite{MIMORADA}\nocite{MIMORADMUTA} finite-length transmit sequences
with impulse like aperiodic auto-correlation function is a necessity.
However, in real life, along with good correlation property, the aforementioned
applications also pose different constraints on the transmit sequence
like the power and spectral (range of operating frequencies) constraints.
The power constraint is mainly due to the limited budget of transmitter
power available in the system. Hence, the design of phase only sequences
of arbitrary lengths having unit magnitude and impulse like aperiodic
auto-correlation function is always desired \cite{_Benedeto_Phase_Magz},
\cite{Stoica_codes_Magz}, \cite{7_probingwaveformrxsyn_robertHeStoica}. 

Earlier, design of phase only sequences is done mainly by algebraic
approaches and some of the sequences designed through algebraic approaches
are Barker sequence \cite{barker_comm}, \cite{12binary}, Frank sequence
\cite{13_Polyphasecode_frank}, Golomb sequence \cite{15_polyphaseseq_Zhang},
Chu sequence \cite{15_polyphaseseq_Zhang}, \cite{3_Radarsignal_Levanon}
and P4 sequence \cite{Impulse_period}, \cite{Set_period}. But all
the above-mentioned sequences exist only for the shorter lengths and
have limited degrees of freedom. Hence, algebraic approaches are not
viable to generate sequences of large lengths. To overcome this issue,
recently, computational approaches \cite{7_probingwaveformrxsyn_robertHeStoica},
\cite{CompBinary}, \cite{CompDes}, \cite{BinRad}, came into existence
and enabled a way to design sequences of arbitrary lengths at a slighter
computational cost. Some of the developed computational approaches
are the stochastic search methods \cite{GSalg}, exhaustive search
methods \cite{11_binaryseq}, which are heuristic in nature with no
guarantee for convergence to a stationary point of ISL function. To
overcome all such issues, very recently several optimization methods
\cite{Frame_MM}, \cite{Opt_Rad_comm}, \cite{Opt_Activesen}, \cite{EffRad}
came into the existence and some of the approaches are CAN \cite{16_CAN},
MISL \cite{17_MISL}, ISL-NEW \cite{20_fast_alg_waveform_LI}, MM-Corr
\cite{18_MMcorr}, ADMM approach \cite{19_ADMM}, MWISL, MWISL-Diag,
MM-PSL \cite{21_MM_PSL}, and CPM \cite{CPM_kehrodi}- a detailed
review of some of the methods will follow shortly. The following mathematical
notations are used hereafter: boldface lowercase letters denote column
vectors, boldface uppercase letters denote matrices and italics denote
scalars. The superscripts $()^{*},()^{T},()^{H}$ denote complex conjugate,
transpose, and conjugate transpose, respectively. $\text{Tr}()$ denote
the trace of a matrix. $z_{m}$ denote the $m^{th}$ element of a
vector $\boldsymbol{\boldsymbol{z}}$. $\text{Re}(.)$ and $\text{Im}(.)$
denote the real and imaginary parts, respectively. $\boldsymbol{\boldsymbol{I}}_{a}$
denote the $a\times a$ identity matrix. $||.||_{2}$ denote the $l_{2}$
norm. $\text{vec}(\boldsymbol{\boldsymbol{S}})$ is a column vector
that consists of all the columns of a matrix-$\boldsymbol{\boldsymbol{S}}$
stacked. $\left|.\right|^{2}$ denote the absolute squared value.
$\text{Diag}(\boldsymbol{\boldsymbol{z}})$ is a diagonal matrix formed
with a vector $\boldsymbol{\boldsymbol{z}}$ as its diagonal. $\mathbb{R}$
and $\mathbb{C}$ represent the real and complex fields. $\lambda_{\text{max}}(\boldsymbol{\boldsymbol{\boldsymbol{R}}})$
denote the maximum eigenvalue of $\boldsymbol{\boldsymbol{\boldsymbol{R}}}$.
$\nabla g(.)$ denote the gradient of a function $g(.).$ $\boldsymbol{\boldsymbol{\boldsymbol{b}}}_{1:N}$
represents the first $N$ elements of a vector $\boldsymbol{\boldsymbol{\boldsymbol{b}}}$.

\subsection*{A. SIGNAL MODEL AND PROBLEM FORMULATION}

Let $\left\{ z_{n}\right\} _{n=1}^{P}$ be a phase only sequence of
length $\text{\textquoteleft}P\text{\textquoteright}$ to be designed.
The $m^{th}$ element of a sequence is denoted as $e^{j\varphi(m)}$,
where $\varphi(m)$ is an arbitrary phase angle that varies between
$0$ and $2\pi$ radians. The aperiodic auto-correlation function
of a sequence $\left\{ z_{n}\right\} _{n=1}^{P}$ at any lag $\text{\textquoteleft}l\lyxmathsym{\textquoteright}$
is defined as:

\begin{equation}
r(l)=\sum_{n=1}^{P-l}z_{n+l}z_{n}^{*}=r^{*}(-l),\hspace{1em}l=0,....,P-1.\label{eq:acor}
\end{equation}

The Integrated Side-lobe Level (ISL) metric, which is a direct measure
of the designed sequence is defined as:

\begin{equation}
\text{ISL}=\sum_{l=1}^{P-1}|r(l)|^{2}.\label{eq:isl}
\end{equation}

The Peak Side-lobe Level (PSL) metric is defined as:

\begin{equation}
\text{PSL}=\text{max}\left\{ |r(l)|\right\} _{l=1}^{P-1}\label{eq:PSL}
\end{equation}

So, the problem to design a phase only sequence that minimizes the
ISL metric is formulated as:

\begin{equation}
\begin{aligned} & \underset{\boldsymbol{\boldsymbol{\boldsymbol{z}}}}{\text{\text{minimize}}} &  & \text{ISL}=\sum_{l=1}^{P-1}|r(l)|^{2}\\
 & \text{subject to} &  & |z_{n}|=1,\;n=1,...,P.
\end{aligned}
\label{prob}
\end{equation}

where $\boldsymbol{\boldsymbol{z}}=[z_{1}z_{2}....z_{P}]_{1\times P}^{T}$.

Apart from the unimodular constraint, there are interests in imposing
constraints like binary constraint \cite{BinRad}, \cite{CompBinary},
\cite{EFFBinSol}, spectral constraint \cite{SpectralShaping}, \cite{fan_psl},
\cite{minmax_logexp}, similarity constraint \cite{UnisimiCons},
\cite{RADARSimi}, Peak to Average Power Ratio (PAPR) constraint \cite{MIMO_channel_est}
to name a few. 

In the next subsection, we will discuss the general framework of majorization-minimization,
which would play a central role in the development of our algorithm.

\subsection*{B. Majorization-Minimization Method}

Majorization-Minimization (MM) is a two-step technique, which is used
to solve the hard (non-convex or even convex) problems very efficiently
\cite{23_Tutorial_MM}, \cite{22_MM_prabhubabu}. The first step of
the MM method is to construct a majorization (upper bound) function
$u()$ to the original objective function $g()$ at any point $\boldsymbol{\boldsymbol{\boldsymbol{z}}}^{k}$
($\boldsymbol{\boldsymbol{\boldsymbol{z}}}$ at $k^{th}$ iteration)
and then second step is to minimize the upper-bound function $u()$
to generate a next update $\boldsymbol{\boldsymbol{\boldsymbol{z}}}^{k+1}$.
So, at every newly generated point, the above mentioned two steps
will be applied repeatedly until it reaches the optimum minimum point
of an original function $g()$. For any given problem, the construction
of a majorization function is not unique and for the same problem,
different types of majorization functions will exist. So, the performance
will depend solely on the chosen majorization function and the different
ways to construct a majorization function are shown in \cite{22_MM_prabhubabu},
\cite{Frame_MM}.

The majorization function $u(\boldsymbol{\boldsymbol{\boldsymbol{z}}}|\boldsymbol{\boldsymbol{\boldsymbol{z}}}^{k})$,
which is constructed in the first step of the MM method has to satisfy
the following properties:

\begin{equation}
u(\boldsymbol{\boldsymbol{\boldsymbol{z}}}^{k}|\boldsymbol{\boldsymbol{\boldsymbol{z}}}^{k})=g(\boldsymbol{\boldsymbol{\boldsymbol{z}}}^{k}),\quad\forall\boldsymbol{\boldsymbol{z}}\in\boldsymbol{\boldsymbol{\boldsymbol{Z}}}.\label{eq:5-1}
\end{equation}

\begin{equation}
u(\boldsymbol{\boldsymbol{z}}|\boldsymbol{\boldsymbol{\boldsymbol{z}}}^{k})\geq g(\boldsymbol{\boldsymbol{z}}),\quad\forall\boldsymbol{\boldsymbol{z}}\in\boldsymbol{\boldsymbol{\boldsymbol{Z}}}.\label{eq:6-1}
\end{equation}

where $\boldsymbol{\boldsymbol{\boldsymbol{Z}}}$ is the set consists
of all the possible values of $\boldsymbol{\boldsymbol{z}}$. As the
MM technique is an iterative process, it will generate the sequence
of points $\left\{ \boldsymbol{\boldsymbol{z}}\right\} =\boldsymbol{\boldsymbol{\boldsymbol{z}}}^{1},\boldsymbol{\boldsymbol{\boldsymbol{z}}}^{2},\boldsymbol{\boldsymbol{\boldsymbol{z}}}^{3},.....,\boldsymbol{\boldsymbol{\boldsymbol{z}}}^{m}$
according to the following update rule:

\begin{equation}
\boldsymbol{\boldsymbol{\boldsymbol{z}}}^{k+1}\triangleq\text{arg}\min_{\boldsymbol{\boldsymbol{z}\in\boldsymbol{\boldsymbol{\boldsymbol{Z}}}}}u(\boldsymbol{\boldsymbol{z}}|\boldsymbol{\boldsymbol{\boldsymbol{z}}}^{k}).\label{eq:7-1}
\end{equation}

The cost function value evaluated at every point generated by (\ref{eq:7-1})
will satisfy the descent property, i.e.
\begin{equation}
g(\boldsymbol{\boldsymbol{\boldsymbol{z}}}^{k+1})\leq u(\boldsymbol{\boldsymbol{\boldsymbol{z}}}^{k+1}|\boldsymbol{\boldsymbol{\boldsymbol{z}}}^{k})\leq u(\boldsymbol{\boldsymbol{\boldsymbol{z}}}^{k}|\boldsymbol{\boldsymbol{\boldsymbol{z}}}^{k})=g(\boldsymbol{\boldsymbol{\boldsymbol{z}}}^{k}).\label{eq:MM-1}
\end{equation}

\subsection*{C. Related work and our Contributions}

The existing algorithms which are developed by solving the same ISL
minimization problem (\ref{prob}) are CAN \cite{16_CAN}, MISL \cite{17_MISL}
, ISL-NEW \cite{20_fast_alg_waveform_LI}, MM-Corr \cite{18_MMcorr},
ADMM approach \cite{19_ADMM}, MWISL, MWISL-Diag \cite{21_MM_PSL},
CPM \cite{CPM_kehrodi}. In the following, we will discuss them briefly
and highlight their potentials and drawbacks. 

Stoica et.al proposed the CAN algorithm \cite{16_CAN}, which works
on the principle of alternating minimization technique. They solved
the problem by transforming the objective function in (\ref{prob})
to the frequency domain as:

\begin{equation}
\sum_{l=1}^{P-1}\mid r(l)\mid^{2}=\frac{1}{4P}\sum_{a=1}^{2P}\Biggl[\Biggl|\sum_{n=1}^{P}z_{n}e^{-j\omega_{a}n}\Biggr|^{2}-P\Biggr]^{2},\label{eq:lag_freq}
\end{equation}

where $\omega_{a}=\frac{2\pi}{2P}a,\:a=1,...,2P.$ are the Fourier
grid frequencies. Then the problem (\ref{prob}) is converted into:
\begin{equation}
\begin{aligned} & \underset{\boldsymbol{\boldsymbol{\boldsymbol{z}}}}{\text{\text{minimize}}} &  & \frac{1}{4P}\sum_{a=1}^{2P}\Biggl[\Biggl|\sum_{n=1}^{P}z_{n}e^{-j\omega_{a}n}\Biggr|^{2}-P\Biggr]^{2}\\
 & \text{subject to} &  & |z_{n}|=1,\:n=1,...,P.
\end{aligned}
\label{eq:canprob}
\end{equation}

The cost function of the problem in (\ref{eq:canprob}) is quartic
in $\boldsymbol{\boldsymbol{\boldsymbol{z}}}$ and it is very hard
to solve further. So, instead of solving (\ref{eq:canprob}) directly,
they solved an approximate problem, which is quadratic in $\boldsymbol{\boldsymbol{\boldsymbol{z}}}$
as shown below:

\begin{equation}
\begin{aligned} & \underset{\boldsymbol{\boldsymbol{\boldsymbol{z}}},\phi_{a}}{\text{\text{minimize}}} &  & \sum_{a=1}^{2P}\Biggl[\Biggl|\sum_{n=1}^{P}z_{n}e^{-j\omega_{a}n}-\sqrt{P}e^{j\phi_{a}}\Biggr|^{2}\Biggr]\\
 & \text{subject to} &  & |z_{n}|=1,\:n=1,...,P,
\end{aligned}
\label{eq:can2}
\end{equation}

where $\phi_{a},$ $a=1,2,...,2P$. are the auxiliary phase variables.
The resulting problem can be rewritten more compactly as follows:

\begin{equation}
\begin{aligned} & \underset{\boldsymbol{\boldsymbol{z}},\boldsymbol{\boldsymbol{y}}}{\text{\text{minimize}}} &  & \Biggl\Vert\boldsymbol{\boldsymbol{\hat{E}}}^{H}\boldsymbol{\hat{\boldsymbol{z}}}-\sqrt{P}\boldsymbol{\boldsymbol{y}}\Biggr\Vert^{2}\\
 & \text{subject to} &  & |z_{n}|=1,n=1,...,P
\end{aligned}
\label{eq:can3}
\end{equation}

where $\boldsymbol{\boldsymbol{\hat{E}}}\triangleq[\boldsymbol{\boldsymbol{e}}_{1},....,\boldsymbol{\boldsymbol{e}}_{2P}]$
be a $2P\times2P$ matrix with $\boldsymbol{\boldsymbol{e}}_{a}\triangleq[e^{j\omega_{a}(1)},e^{j\omega_{a}(2)},...,e^{j\omega_{a}(2P)}]^{T}$,
$\boldsymbol{\hat{\boldsymbol{z}}}\triangleq[z_{1},z_{2},....z_{P},0,...,0]_{1\times2P}^{T}$
and $\boldsymbol{\boldsymbol{y}}\triangleq[e^{j\phi_{1}},...,e^{j\phi_{2P}}]^{T}$.
They solved the problem in (\ref{eq:can3}) by alternatively minimizing
between the variables $\boldsymbol{\boldsymbol{z}}$ and $\boldsymbol{\boldsymbol{y}}$.
For a given $\boldsymbol{\boldsymbol{z}}$, minimization of (\ref{eq:can3})
with respeect to $\boldsymbol{\boldsymbol{y}}$ is given by:

\begin{equation}
\boldsymbol{y}=\frac{\boldsymbol{\boldsymbol{\boldsymbol{v}}}}{||\boldsymbol{\boldsymbol{\boldsymbol{v}}}||_{2}},\label{eq:pfiupdate}
\end{equation}

where $\boldsymbol{\boldsymbol{v}}\triangleq\boldsymbol{\hat{\boldsymbol{E}}}^{H}\boldsymbol{\boldsymbol{\hat{\boldsymbol{z}}}}$
($\boldsymbol{\hat{\boldsymbol{E}}}^{H}$ is a $2N\times2N$ FFT matrix
) and for a fixed $\boldsymbol{\boldsymbol{y}}$, minimizer over $\boldsymbol{\boldsymbol{z}}$
would be:

\begin{equation}
\boldsymbol{\boldsymbol{z}}=\frac{\boldsymbol{\boldsymbol{\boldsymbol{b}}}}{||\boldsymbol{\boldsymbol{\boldsymbol{b}}}||_{2}},\label{eq:yupdate}
\end{equation}

where $\boldsymbol{\boldsymbol{b}}\triangleq\boldsymbol{\boldsymbol{\hat{E}}}\boldsymbol{\boldsymbol{y}}$
($\boldsymbol{\hat{\boldsymbol{E}}}$ is a $2N\times2N$ IFFT matrix
). The pseudocode of the CAN algorithm is summarized in Algorithm
1.

\begin{algorithm}[h]
\textbf{Require}: sequence length $\text{\textquoteleft}P\text{\textquoteright}$

1: set $k=0$, initialize $\boldsymbol{\boldsymbol{z}}^{0}$

2:\textbf{ repeat}

3: $\hphantom{n}$$\boldsymbol{\boldsymbol{v}}=\boldsymbol{\hat{E}}^{H}\boldsymbol{\boldsymbol{\hat{\boldsymbol{z}}}}^{k}$

4: $\hphantom{n}$$\boldsymbol{y}=\frac{\boldsymbol{\boldsymbol{\boldsymbol{v}}}}{||\boldsymbol{\boldsymbol{\boldsymbol{v}}}||_{2}}$

5: $\hphantom{n}$$\boldsymbol{\boldsymbol{b}}=\boldsymbol{\hat{E}}\boldsymbol{\boldsymbol{y}}$

6: $\hphantom{n}$$\boldsymbol{\boldsymbol{z}}^{k+1}=\frac{\boldsymbol{\boldsymbol{\boldsymbol{b}}}_{1:P}}{||\boldsymbol{\boldsymbol{\boldsymbol{b}}}_{1:P}||_{2}}$

7: $\hphantom{n}$$k$$\leftarrow$$k+1$

8:\textbf{ until} convergence

\caption{:The CAN algorithm proposed in \cite{16_CAN}}
\end{algorithm}

We would like to point out that, instead of solving the original problem
(\ref{prob}), the CAN algorithm had solved an approximately equivalent
problem (\ref{eq:can2}). So, there is no guarantee for an obtained
minimum of (\ref{eq:can2}) is also a minimum of the original problem
in (\ref{prob}).

To overcome this issue, Song et.al. proposed the MISL algorithm \cite{17_MISL}
by solving the original problem (\ref{prob}) directly via the MM
approach. So, from (\ref{eq:canprob}) we have,

\[
\begin{aligned} & \underset{\boldsymbol{\boldsymbol{\boldsymbol{z}}}}{\text{\text{minimize}}} &  & \frac{1}{4P}\sum_{a=1}^{2P}\Biggl[\Biggl|\sum_{n=1}^{P}z_{n}e^{-j\omega_{a}n}\Biggr|^{2}-P\Biggr]^{2}\\
 & \text{subject to} &  & |z_{n}|=1,\:n=1,...,P.
\end{aligned}
\]

By expanding the cost function and ignoring the constant and multiplication
terms, the above problem can be rewritten more compactly as:

\begin{equation}
\begin{aligned} & \underset{\boldsymbol{\boldsymbol{\boldsymbol{z}}}}{\text{\text{minimize}}} &  & \sum_{a=1}^{2P}\Biggl[\boldsymbol{\boldsymbol{\boldsymbol{e}}_{a}}^{H}\boldsymbol{\boldsymbol{\boldsymbol{z}}}\boldsymbol{\boldsymbol{\boldsymbol{z}}}^{H}\boldsymbol{\boldsymbol{\boldsymbol{e}}_{a}}\Biggr]^{2}\\
 & \text{subject to} &  & |z_{n}|=1,\:n=1,...,P.
\end{aligned}
\label{eq:MISL1}
\end{equation}

In terms of $\boldsymbol{\boldsymbol{z}}$, the problem in (\ref{eq:MISL1})
is quartic and very hard to solve further. So, by defining $\boldsymbol{\boldsymbol{\boldsymbol{Z}}}=\boldsymbol{\boldsymbol{\boldsymbol{z}}}\boldsymbol{\boldsymbol{z}}^{H}$
and $\boldsymbol{\boldsymbol{\boldsymbol{C}}_{a}}=\boldsymbol{\boldsymbol{\boldsymbol{e}}_{a}}\boldsymbol{\boldsymbol{\boldsymbol{\boldsymbol{e}}_{a}}}^{H}$
, problem in (\ref{eq:MISL1}) can be rewritten as:

\begin{equation}
\begin{aligned} & \underset{\boldsymbol{\boldsymbol{\boldsymbol{z}}},\boldsymbol{\boldsymbol{\boldsymbol{Z}}}}{\text{\text{minimize}}} &  & \text{vec}(\boldsymbol{\boldsymbol{\boldsymbol{Z}}})^{H}\boldsymbol{\boldsymbol{\boldsymbol{\Phi}}}\text{vec}(\boldsymbol{\boldsymbol{\boldsymbol{Z}}})\\
 & \text{subject to} &  & |z_{n}|=1,\:n=1,...,P,\\
 &  &  & \boldsymbol{\boldsymbol{\boldsymbol{Z}}}=\boldsymbol{\boldsymbol{\boldsymbol{z}}}\boldsymbol{\boldsymbol{z}^{H}},
\end{aligned}
\label{eq:MISL2}
\end{equation}

where $\boldsymbol{\boldsymbol{\boldsymbol{\Phi}}}=\stackrel[a=1]{2P}{\sum}\text{vec}(\boldsymbol{\boldsymbol{\boldsymbol{C}}_{a}})\text{vec}(\boldsymbol{\boldsymbol{\boldsymbol{C}}_{a}})^{H}$.
The cost function in (\ref{eq:MISL2}) is quadratic in $\boldsymbol{\boldsymbol{\boldsymbol{Z}}}$.
So, they constructed a majorization function for it by using second-order
Taylor series method \cite{22_MM_prabhubabu}, \cite{Frame_MM}, and
by neglecting the constant terms, the surrogate problem can be rewritten
more compactly as:

\begin{equation}
\begin{aligned} & \underset{\boldsymbol{\boldsymbol{\boldsymbol{z}}}}{\text{\text{minimize}}} &  & \boldsymbol{\boldsymbol{\boldsymbol{z}}^{H}}\Biggl[\boldsymbol{\boldsymbol{\hat{E}}}\text{Diag}(\boldsymbol{\boldsymbol{\boldsymbol{b}}^{k}})\boldsymbol{\hat{E}}^{H}-2P^{2}\boldsymbol{\boldsymbol{\boldsymbol{z}}^{k}}\boldsymbol{(\boldsymbol{\boldsymbol{z}^{k}})^{H}}\Biggr]\boldsymbol{\boldsymbol{\boldsymbol{z}}}\\
 & \text{subject to} &  & |z_{n}|=1,\:n=1,...,P,
\end{aligned}
\label{eq:MISL3}
\end{equation}

where $\boldsymbol{\boldsymbol{\boldsymbol{b}}^{k}}=\Bigl|\boldsymbol{\boldsymbol{\hat{E}}}^{H}\boldsymbol{\boldsymbol{z}}^{k}\Bigr|.$
The resultant problem in (\ref{eq:MISL3}) is quadratic in $\boldsymbol{\boldsymbol{z}}$,
and they have majorized the cost function in the above problem once
again as mentioned above, to obtain a simple closed-form solution.
After majorizing for the second time and by ignoring the constant
terms, the final surrogate minimization problem becomes:

\begin{equation}
\begin{aligned} & \underset{\boldsymbol{\boldsymbol{\boldsymbol{z}}}}{\text{\text{minimize}}} &  & \text{Re}\Biggl(\boldsymbol{\boldsymbol{\boldsymbol{z}}^{H}}\Biggl[\boldsymbol{\tilde{\boldsymbol{C}}}-2P^{2}\boldsymbol{\boldsymbol{\boldsymbol{z}}^{k}}\boldsymbol{(\boldsymbol{\boldsymbol{z}^{k}})^{H}}\Biggr]\boldsymbol{\boldsymbol{\boldsymbol{z}}^{k}}\Biggr)\\
 & \text{subject to} &  & |z_{n}|=1,\:n=1,...,P,
\end{aligned}
\label{eq:MISL4}
\end{equation}

where $\boldsymbol{\tilde{\boldsymbol{C}}}=\boldsymbol{\hat{E}}\Biggl(\text{Diag}(\boldsymbol{\boldsymbol{\boldsymbol{b}}^{2k}})-b_{max}^{k}\boldsymbol{\boldsymbol{I}}\Biggr)\boldsymbol{\boldsymbol{\hat{E}}}^{H}$
and $b_{max}^{k}=\underset{a}{\text{max}}\bigl\{(b_{a}^{k})^{2},\:a=1,2,.,2P\bigr\}$.
Problem in (\ref{eq:MISL4}) can be rewritten more compactly as:

\begin{equation}
\begin{aligned} & \underset{\boldsymbol{\boldsymbol{\boldsymbol{z}}}}{\text{\text{minimize}}} &  & \parallel\boldsymbol{\boldsymbol{\boldsymbol{z}}}-\boldsymbol{\boldsymbol{\boldsymbol{d}}}\parallel_{2}^{2}\\
 & \text{subject to} &  & |z_{n}|=1,\:n=1,...,P,
\end{aligned}
\label{eq:misl5}
\end{equation}

where $\boldsymbol{\boldsymbol{\boldsymbol{d}}}=-\boldsymbol{\hat{E}}\Biggl(\text{Diag}(\boldsymbol{\boldsymbol{\boldsymbol{b}}^{2k}})-b_{max}^{k}\boldsymbol{\boldsymbol{I}}-N^{2}\boldsymbol{\boldsymbol{I}}\Biggr)\boldsymbol{\boldsymbol{\hat{E}}}^{H}\boldsymbol{\boldsymbol{\boldsymbol{z}}^{k}}$.
The problem in (\ref{eq:misl5}) has a closed-form solution:

\begin{equation}
\boldsymbol{\boldsymbol{\boldsymbol{z}}}=\frac{\boldsymbol{\boldsymbol{\boldsymbol{d}}}}{||\boldsymbol{\boldsymbol{\boldsymbol{d}}}||_{2}}.\label{eq:misl6}
\end{equation}

The pseudocode of the MISL algorithm is summarized in Algorithm 2.

\begin{algorithm}[h]
\textbf{Require}: sequence length $\text{\textquoteleft}P\text{\textquoteright}$

1: set $k=0$, initialize $\boldsymbol{\boldsymbol{z}}^{k}$

2:\textbf{ repeat}

3: $\hphantom{n}$$\boldsymbol{\boldsymbol{b}}^{k}=\Bigl|\boldsymbol{\boldsymbol{\hat{E}}}^{H}\boldsymbol{\boldsymbol{z}}^{k}\Bigr|$

4: $\hphantom{n}$$b_{\text{max}}^{k}=\underset{a}{\text{max}}\bigl\{(b_{a}^{k})^{2},\:a=1,..,2P\bigr\}$

5: $\hphantom{n}$$\boldsymbol{\boldsymbol{d}}=-\boldsymbol{\boldsymbol{\hat{E}}}\Biggl(\text{Diag}\Bigl(\boldsymbol{\boldsymbol{b}}^{2k}\Bigr)-b_{\text{max}}^{k}\boldsymbol{\boldsymbol{I}}-N^{2}\boldsymbol{\boldsymbol{I}}\Biggr)\boldsymbol{\boldsymbol{\hat{E}}}^{H}\boldsymbol{\boldsymbol{z}}^{k}$

6: $\hphantom{n}$$\boldsymbol{\boldsymbol{\boldsymbol{z}}}^{k+1}=\frac{\boldsymbol{\boldsymbol{\boldsymbol{d}}}_{1:P}}{||\boldsymbol{\boldsymbol{\boldsymbol{d}}}_{1:P}||_{2}}$

7: $\hphantom{n}$$k$$\leftarrow$$k+1$

8:\textbf{ until }convergence\caption{:The MISL algorithm proposed in \cite{17_MISL}}
\end{algorithm}

Compared to the CAN algorithm, the MISL algorithm solves the original
problem in (\ref{prob}). So, there is an assurance of obtaining an
original optimum minimum point. But, the MISL algorithm faces a drawback
of slower convergence due to twice the majorization of the original
objective function. To deal with the convergence issue, they have
proposed acceleration schemes to accelerate the MISL algorithm.

In \cite{20_fast_alg_waveform_LI}, Y. Li et.al proposed an algorithm
named ISL-NEW using the MM method to design sequence-set. By particularizing
it for single sequence, we observe that the only difference between
the MISL and ISL-NEW algorithms is in the way they arrive at their
majorizing functions. After majorizing the objective function in (\ref{eq:MISL2})
and removing the constant terms, the final surrogate problem they
solve is given by:

\begin{equation}
\begin{aligned} & \underset{\boldsymbol{\boldsymbol{\boldsymbol{z}}}}{\text{\text{minimize}}} &  & \boldsymbol{\boldsymbol{\boldsymbol{z}}^{H}}\Biggl[\boldsymbol{\hat{E}}\text{Diag}(\boldsymbol{\boldsymbol{\boldsymbol{b}}^{k}})\boldsymbol{\boldsymbol{\hat{E}}}^{H}-P^{2}\boldsymbol{\boldsymbol{\boldsymbol{z}}^{k}}\boldsymbol{(\boldsymbol{\boldsymbol{z}^{k}})^{H}}\Biggr]\boldsymbol{\boldsymbol{\boldsymbol{z}}}\\
 & \text{subject to} &  & |z_{n}|=1,\:n=1,...,P.
\end{aligned}
\label{eq:MISL3-1}
\end{equation}

The resultant problem in (\ref{eq:MISL3-1}) is quadratic in $\boldsymbol{\boldsymbol{z}}$.
So, they majorized the cost function in (\ref{eq:MISL3-1}) once again
and arrive at the following problem:

\begin{equation}
\begin{aligned} & \underset{\boldsymbol{\boldsymbol{\boldsymbol{z}}}}{\text{\text{minimize}}} &  & \text{Re}\Biggl(\boldsymbol{\boldsymbol{\boldsymbol{z}}^{H}}\Biggl[\boldsymbol{\bar{C}}-P^{2}\boldsymbol{\boldsymbol{\boldsymbol{z}}^{k}}\boldsymbol{(\boldsymbol{\boldsymbol{z}^{k}})^{H}}\Biggr]\boldsymbol{\boldsymbol{\boldsymbol{z}}^{k}}\Biggr)\\
 & \text{subject to} &  & |z_{n}|=1,\:n=1,...,P,
\end{aligned}
\label{eq:MISL4-1}
\end{equation}

where $\boldsymbol{\bar{C}}=\boldsymbol{\hat{E}}\Biggl(\text{Diag}(\boldsymbol{\boldsymbol{\boldsymbol{b}}^{2k}})-0.5b_{max}^{k}\boldsymbol{\boldsymbol{I}}\Biggr)\boldsymbol{\boldsymbol{\hat{E}}}^{H}$
. The problem in (\ref{eq:MISL4-1}) can be rewritten as:

\begin{equation}
\begin{aligned} & \underset{\boldsymbol{\boldsymbol{\boldsymbol{z}}}}{\text{\text{minimize}}} &  & \parallel\boldsymbol{\boldsymbol{\boldsymbol{z}}}-\boldsymbol{\boldsymbol{\hat{d}}}\parallel_{2}^{2}\\
 & \text{subject to} &  & |z_{n}|=1,\:n=1,...,P,
\end{aligned}
\label{eq:misl5-1}
\end{equation}

where $\boldsymbol{\boldsymbol{\hat{d}}}=-\boldsymbol{\hat{E}}\Biggl(\text{Diag}(\boldsymbol{\boldsymbol{\boldsymbol{b}}^{2k}})-0.5b_{max}^{k}\boldsymbol{\boldsymbol{I}}-0.5N^{2}\boldsymbol{\boldsymbol{I}}\Biggr)\boldsymbol{\boldsymbol{\hat{E}}}^{H}\boldsymbol{\boldsymbol{\boldsymbol{z}}^{k}}$.
The problem in (\ref{eq:misl5-1}) has a closed-form solution

\begin{equation}
\boldsymbol{\boldsymbol{\boldsymbol{z}}}=\frac{\boldsymbol{\boldsymbol{\boldsymbol{\hat{d}}}}}{||\boldsymbol{\boldsymbol{\hat{d}}}||_{2}}.\label{eq:misl6-1}
\end{equation}

The pseudocode of the ISL-NEW algorithm is summarized in Algorithm
3.

\begin{algorithm}[h]
\textbf{Require}: sequence length $\text{\textquoteleft}P\text{\textquoteright}$

1: set $k=0$, initialize $\boldsymbol{\boldsymbol{z}}^{k}$

2:\textbf{ repeat}

3: $\hphantom{n}$$\boldsymbol{\boldsymbol{b}}^{k}=\Bigl|\boldsymbol{\boldsymbol{\hat{E}}}^{H}\boldsymbol{\boldsymbol{z}}^{k}\Bigr|$

4: $\hphantom{n}$$b_{\text{max}}^{k}=\underset{a}{\text{max}}\bigl\{(b_{a}^{k})^{2}:a=1,..,2P\bigr\}$

5: $\hphantom{n}$$\boldsymbol{\boldsymbol{\hat{d}}}=-\boldsymbol{\boldsymbol{\hat{E}}}\Biggl(\text{Diag}\Bigl(\boldsymbol{\boldsymbol{b}}^{2k}\Bigr)-0.5b_{\text{max}}^{k}\boldsymbol{\boldsymbol{I}}-0.5N^{2}\boldsymbol{\boldsymbol{I}}\Biggr)\boldsymbol{\boldsymbol{\hat{E}}}^{H}\boldsymbol{\boldsymbol{z}}^{k}$

6: $\hphantom{n}$$\boldsymbol{\boldsymbol{\boldsymbol{z}}}^{k+1}=\frac{\boldsymbol{\boldsymbol{\boldsymbol{\hat{d}}}}_{1:P}}{||\boldsymbol{\boldsymbol{\boldsymbol{\hat{d}}}}_{1:P}||_{2}}$

7: $\hphantom{n}$$k$$\leftarrow$$k+1$

8:\textbf{ until }convergence\caption{:The ISL-NEW algorithm proposed in \cite{20_fast_alg_waveform_LI}}
\end{algorithm}

Y. Li et.al has also solved the problem (\ref{prob}) directly and
concluded it as a fast algorithm in terms of the convergence. However,
due to similarity in the update step of ISL-NEW and MISL (with very
little difference), ISL-NEW also suffers from slow convergence and
they have also proposed acceleration schemes to accelerate the ISL-NEW
algorithm. The above mentioned three algorithms CAN, MISL and ISL-NEW
can be implemented via FFT and IFFT operations. Hence, they are computationally
efficient for generating sequences of large lengths.

In \cite{18_MMcorr}, J. Song et.al proposed an algorithm named as
MM-Corr to design the sequence set using the MM method. In \cite{19_ADMM},
J.Liang et.al proposed a new algorithm by solving the approximately
equivalent problem to problem (\ref{prob}) (i.e, same as CAN algorithm)
by using the ADMM method and they concluded that its performance is
worse than the MISL algorithm in terms of the PSL metric value in
its aperiodic autocorrelation function. In \cite{21_MM_PSL}, J.song
et.al proposed three different algorithms named MWISL, MWISL-Diag,
and MM-PSL by using the MM method. We observe that out of three algorithms
MWISL and MWISL-Diag are variants of the MISL algorithm \cite{17_MISL}
and MM-PSL algorithm is derived by solving the $l_{p}$-norm, $2<p<\infty$,
which is different from the ISL metric. In \cite{CPM_kehrodi}, Mohammad
et.al has proposed an algorithm named as CPM based on the coordinate
descent framework and concluded that CPM performs well only in the
case of binary and finite discrete phase constraints. Some more algorithms,
which are derived based on different metrics like PSL \cite{fan_psl},
\cite{21_MM_PSL}, \cite{PSl_MIMO}, ambiguity function shaping \cite{dopRobdes},
\cite{AmbiguUnimod}, \cite{Local_Ambig_fun}, \cite{aubry}, SINR
\cite{RADAR_SINR}, beam pattern synthesis \cite{fan_mimo}, \cite{MIMO_lp},
are used to design sequences. 

So, the summary of the related literature is as follows:
\begin{itemize}
\item CAN algorithm has solved the approximate problem in (\ref{eq:can2})
and there is no guarantee for an obtained minimum to be also the minimum
of the original problem in (\ref{prob}).
\item Even though the MISL and ISL-NEW algorithms has solved the original
problem in (\ref{prob}), they face a drawback of slower convergence
due to two times the majorization of the original objective function.
\item In comparison to the CAN and MISL algorithms, the authors in \cite{20_fast_alg_waveform_LI}
claimed that the ISL-NEW algorithm is fast but it is only a marginal
improvement. 
\item ADMM algorithm solves the approximate problem (same as CAN algorithm)
and it is a non-monotonic and does not minimize the ISL function.
\item The CPM algorithm is derived based on the coordinate descent method,
as the length of the sequence increases its computational complexity
will also increase and convergence to a minimizer will also get slower.
\end{itemize}
As all the above mentioned state-of-the-art algorithms have either
slower convergence or do not solve the original ISL minimization problem.
This motivated us to solve the original ISL minimization problem (\ref{prob})
with a faster algorithm and we named our algorithm as FISL (Faster
ISL minimization algorithm).

The major contributions of this paper are as follows:
\begin{itemize}
\item An algorithm based on the MM framework is proposed, to design phase
only sequences of arbitrary length $P$ by minimizing the ISL metric.
\item To obtain faster convergence speed, we constructed a majorization
function that acts like a tighter global upper bound to the original
ISL function.
\item Through MATLAB simulations we compare different ways of constructing
a majorization function and pick out the best approach to implement
our algorithm.
\item By using FFT and IFFT operations, we show a computationally efficient
way of implementing our proposed algorithm.
\item We prove that the proposed algorithm converges to a stationary point
of a problem in (\ref{prob}).
\item Numerical experiments were conducted to prove that our proposed algorithm
performs better than the state-of-the-art algorithms in terms of the
speed of convergence.
\end{itemize}
The rest of the paper is organized as follows. In section II, we propose
our algorithm and discuss its convergence analysis, computational
\& space complexities. Section III consists of numerical experiments
and finally, section IV concludes the paper.

\section*{\centerline{II.FISL-Faster ISL Minimization Algorithm}}

\subsection*{A. ISL minimization via MM method}

From (\ref{prob}), we have

\[
\begin{aligned} & \underset{\boldsymbol{\boldsymbol{\boldsymbol{z}}}}{\text{\text{minimize}}} &  & \text{ISL}=\sum_{l=1}^{P-1}|r(l)|^{2}.\\
 & \text{subject to} &  & |z_{n}|=1,\;n=1,...,P.
\end{aligned}
\]

During the problem formulation, we considered only the positive lags,
but now we will reframe it to make the problem of interest consists
of both the positive and negative lags along with the zeroth lag (due
to the unimodular property always equal to the length of a sequence
$P$, which is a constant value).

So, the problem of interest becomes as:

\begin{equation}
\begin{aligned} & \underset{\boldsymbol{\boldsymbol{\boldsymbol{z}}}}{\text{\text{minimize}}} &  & g(\boldsymbol{\boldsymbol{z}})=\sum_{l=-(P-1)}^{P-1}|r(l)|^{2}\\
 & \text{subject to} &  & |z_{n}|=1,\;n=1,...,P.
\end{aligned}
\label{eq:mainprob}
\end{equation}

We can write $r(l)=\boldsymbol{\boldsymbol{\boldsymbol{z}}}^{H}\boldsymbol{\boldsymbol{\boldsymbol{W}}}_{l}\boldsymbol{\boldsymbol{\boldsymbol{z}}}$,
where $\boldsymbol{\boldsymbol{\boldsymbol{W}}}_{l}$ is a Toeplitz
matrix of dimension $P\times P$, with entries given by:
\begin{eqnarray}
\boldsymbol{\boldsymbol{\boldsymbol{W}}}_{l} & = & \begin{cases}
1 & ;j-i=l\\
0 & ;else
\end{cases}\label{eq:toep}
\end{eqnarray}

$i,j$ denote the row and column indexes of $\boldsymbol{\boldsymbol{\boldsymbol{W}}}_{l}$
respectively.

So, the objective function of a problem in (\ref{eq:mainprob}) can
be rewritten as $g(\boldsymbol{\boldsymbol{\boldsymbol{z}}})=\boldsymbol{\boldsymbol{\boldsymbol{z}}}^{H}\boldsymbol{\boldsymbol{\boldsymbol{R}}}(\boldsymbol{\boldsymbol{\boldsymbol{z}}})\boldsymbol{\boldsymbol{\boldsymbol{z}}}$,
where 
\begin{equation}
\boldsymbol{\boldsymbol{\boldsymbol{R}}}(\boldsymbol{\boldsymbol{\boldsymbol{z}}})=\sum_{l=1}^{P-1}r^{*}(l)\boldsymbol{\boldsymbol{\boldsymbol{W}}}_{l}+\sum_{l=1}^{P-1}r(l)\boldsymbol{\boldsymbol{\boldsymbol{W}}}_{l}^{H}+Diag(\boldsymbol{\boldsymbol{r}}_{c}).\label{eq:mid}
\end{equation}

where $\boldsymbol{\boldsymbol{r}}_{c}=[r(0),r(0),....,r(0)]_{1\times P}^{T}$.
So, 

\begin{equation}
\boldsymbol{\boldsymbol{\boldsymbol{R}}}(\boldsymbol{\boldsymbol{\boldsymbol{z}}})=\begin{bmatrix}r(0) & r^{*}(1) & . & . & r^{*}(P-2) & r^{*}(P-1)\\
r(1) & r(0) & r^{*}(1) & . & . & r^{*}(P-2)\\
. & r(1) & r(0) & r^{*}(1) & . & .\\
. & . & r(1) & . & . & .\\
r(P-2) & . & . & . & . & r^{*}(1)\\
r(P-1) & r(P-2) & . & . & r(1) & r(0)
\end{bmatrix}\label{eq:midMat}
\end{equation}

is a Hermitian Toeplitz matrix and to implement it, one can find autocorrelation
of $\boldsymbol{\boldsymbol{\boldsymbol{z}}}$ using FFT and IFFT
operations as: 

\begin{equation}
\boldsymbol{\boldsymbol{\boldsymbol{r}}}=\boldsymbol{\boldsymbol{\hat{E}}}\mid\boldsymbol{\boldsymbol{\hat{E}}}^{H}\boldsymbol{\boldsymbol{z}}\mid^{2}.\label{eq:rfft}
\end{equation}

Here $\mid.\mid^{2}$ is element wise operation. Then the problem
of interest (\ref{eq:mainprob}) becomes as:

\begin{equation}
\begin{aligned} & \underset{\boldsymbol{\boldsymbol{\boldsymbol{z}}}}{\text{\text{minimize}}} &  & g(\boldsymbol{\boldsymbol{\boldsymbol{z}}})=\boldsymbol{\boldsymbol{\boldsymbol{z}}}^{H}\boldsymbol{\boldsymbol{\boldsymbol{R}}}(\boldsymbol{\boldsymbol{\boldsymbol{z}}})\boldsymbol{\boldsymbol{\boldsymbol{z}}}\\
 & \text{subject to} &  & |z_{n}|=1,\;n=1,...,P.
\end{aligned}
\label{eq:prob2}
\end{equation}

In the following, we will introduce a lemma which will be useful in
deriving a majorizing function for the objective in (\ref{eq:prob2}).

\textbf{Lemma-1}: Let $f:\mathbb{\mathbb{C}}^{N}\rightarrow\mathbb{R}$
be a continuously twice differentiable function and if $f(\boldsymbol{\boldsymbol{\boldsymbol{x}}})$
has a bounded curvature, then there exists a matrix $\boldsymbol{\boldsymbol{M}}\succeq\nabla^{2}f(\boldsymbol{\boldsymbol{\boldsymbol{x}}})$,
such that by using the second-order Taylor series expansion, at any
fixed point $\boldsymbol{\boldsymbol{x}}^{k}$, $f(\boldsymbol{\boldsymbol{\boldsymbol{x}}})$
can be upper bounded (majorized) as,

\begin{equation}
f(\boldsymbol{\boldsymbol{\boldsymbol{x}}})=f(\boldsymbol{\boldsymbol{\boldsymbol{x}}}^{k})+\nabla f(\boldsymbol{\boldsymbol{\boldsymbol{x}}}^{k})^{H}(\boldsymbol{\boldsymbol{\boldsymbol{x}}}-\boldsymbol{\boldsymbol{\boldsymbol{x}}}^{k})+\frac{1}{2}(\boldsymbol{\boldsymbol{\boldsymbol{x}}}-\boldsymbol{\boldsymbol{\boldsymbol{x}}}^{k})^{H}\nabla^{2}f(\boldsymbol{\boldsymbol{\boldsymbol{x}}}^{k})(\boldsymbol{\boldsymbol{\boldsymbol{x}}}-\boldsymbol{\boldsymbol{\boldsymbol{x}}}^{k})\label{eq:orgeq}
\end{equation}

\begin{equation}
f(\boldsymbol{\boldsymbol{\boldsymbol{x}}})\leq f(\boldsymbol{\boldsymbol{\boldsymbol{x}}}^{k})+\nabla f(\boldsymbol{\boldsymbol{\boldsymbol{x}}}^{k})^{H}(\boldsymbol{\boldsymbol{\boldsymbol{x}}}-\boldsymbol{\boldsymbol{\boldsymbol{x}}}^{k})+\frac{1}{2}(\boldsymbol{\boldsymbol{\boldsymbol{x}}}-\boldsymbol{\boldsymbol{\boldsymbol{x}}}^{k})^{H}\boldsymbol{\boldsymbol{M}}(\boldsymbol{\boldsymbol{\boldsymbol{x}}}-\boldsymbol{\boldsymbol{\boldsymbol{x}}}^{k})\label{eq:surg}
\end{equation}

Proof: The proof can be found in \cite{22_MM_prabhubabu} $\hphantom{nnnnnnnnnnnnnnnnnnnnnnnnnnnnnnnnnnnnnnnnnnnnnnnnn}$$\blacksquare$

So, according to the lemma-1, by using the second-order Taylor series
expansion, at any fixed point $\boldsymbol{\boldsymbol{\boldsymbol{z}}}^{k}$,
the objective function of the problem in (\ref{eq:prob2}) can be
majorized as,

\[
\boldsymbol{\boldsymbol{\boldsymbol{z}}}^{H}\boldsymbol{\boldsymbol{\boldsymbol{R}}(\boldsymbol{\boldsymbol{\boldsymbol{z}}})}\boldsymbol{\boldsymbol{\boldsymbol{z}}}=(\boldsymbol{\boldsymbol{\boldsymbol{z}}}^{k})^{H}\boldsymbol{\boldsymbol{\boldsymbol{R}}}(\boldsymbol{\boldsymbol{\boldsymbol{z}}}^{k})\boldsymbol{\boldsymbol{\boldsymbol{z}}}^{k}+\text{Re}((4\boldsymbol{\boldsymbol{\boldsymbol{R}}}(\boldsymbol{\boldsymbol{\boldsymbol{z}}}^{k})\boldsymbol{\boldsymbol{\boldsymbol{z}}}^{k})^{H}(\boldsymbol{\boldsymbol{\boldsymbol{z}}}-\boldsymbol{\boldsymbol{\boldsymbol{z}}}^{k}))+\frac{1}{2}(\boldsymbol{\boldsymbol{\boldsymbol{z}}}-\boldsymbol{\boldsymbol{\boldsymbol{z}}}^{k})^{H}(8\boldsymbol{\boldsymbol{\boldsymbol{R}}}(\boldsymbol{\boldsymbol{\boldsymbol{z}}}^{k}))(\boldsymbol{\boldsymbol{\boldsymbol{z}}}-\boldsymbol{\boldsymbol{\boldsymbol{z}}}^{k})
\]

\begin{equation}
\boldsymbol{\boldsymbol{\boldsymbol{z}}}^{H}\boldsymbol{\boldsymbol{\boldsymbol{R}}(\boldsymbol{\boldsymbol{\boldsymbol{z}}})}\boldsymbol{\boldsymbol{\boldsymbol{z}}}\leq(\boldsymbol{\boldsymbol{\boldsymbol{z}}}^{k})^{H}\boldsymbol{\boldsymbol{\boldsymbol{R}}}(\boldsymbol{\boldsymbol{\boldsymbol{z}}}^{k})\boldsymbol{\boldsymbol{\boldsymbol{z}}}^{k}+\text{Re}((4\boldsymbol{\boldsymbol{\boldsymbol{R}}}(\boldsymbol{\boldsymbol{\boldsymbol{z}}}^{k})\boldsymbol{\boldsymbol{\boldsymbol{z}}}^{k})^{H}(\boldsymbol{\boldsymbol{\boldsymbol{z}}}-\boldsymbol{\boldsymbol{\boldsymbol{z}}}^{k}))+\frac{1}{2}(\boldsymbol{\boldsymbol{\boldsymbol{z}}}-\boldsymbol{\boldsymbol{\boldsymbol{z}}}^{k})^{H}(\boldsymbol{\boldsymbol{\boldsymbol{M}}})(\boldsymbol{\boldsymbol{z}}-\boldsymbol{\boldsymbol{z}}^{k})\label{eq:st2}
\end{equation}

There are more than one way to construct a matrix $\boldsymbol{\boldsymbol{\boldsymbol{M}}}$,
such that (\ref{eq:st2}) holds, some simple ways would be to choose:

\begin{equation}
\boldsymbol{\boldsymbol{M}}=\text{Tr}(8\boldsymbol{\boldsymbol{\boldsymbol{R}}}(\boldsymbol{\boldsymbol{\boldsymbol{z}}}^{k}))\boldsymbol{\boldsymbol{I}}_{P}=8P^{2}\boldsymbol{\boldsymbol{I}}_{P}.\label{eq:Trace}
\end{equation}

or

\begin{equation}
\boldsymbol{\boldsymbol{M}}=\lambda_{\text{max}}(8\boldsymbol{\boldsymbol{\boldsymbol{R}}}(\boldsymbol{\boldsymbol{\boldsymbol{z}}}^{k}))\boldsymbol{\boldsymbol{I}}_{P}.\label{eq:eig}
\end{equation}

But in practice, for large dimension sequences, calculating the maximum
eigenvalue is a computationally demanding procedure. So, in the following
we try to explore the tighter upper bounds on maximum eigenvalue of
the Hessian matrix.

\textbf{Theorem-1} {[}Theorem 2.1 \cite{Bound_Eig}{]}: Let $\boldsymbol{\boldsymbol{\boldsymbol{A}}}$
be a $P\times P$ matrix with complex entries having real eigenvalues
and let

\begin{equation}
m=\frac{1}{P}\text{Tr}(\boldsymbol{\boldsymbol{\boldsymbol{A}}}),\hphantom{n}s^{2}=(\frac{1}{P}\text{Tr}(\boldsymbol{\boldsymbol{\boldsymbol{A}}}^{2}))-m^{2}\label{eq:trlam}
\end{equation}

Then

\begin{equation}
m-s(P-1)^{1/2}\leq\lambda_{\text{min}}(\boldsymbol{\boldsymbol{\boldsymbol{A}}})\leq m-\frac{s}{(P-1)^{1/2}}\label{eq:mintralam}
\end{equation}

\begin{equation}
m+\frac{s}{(P-1)^{1/2}}\leq\lambda_{\text{max}}(\boldsymbol{\boldsymbol{\boldsymbol{A}}})\leq m+s(P-1)^{1/2}\label{eq:maxtralam}
\end{equation}

So, by using the result from Theorem-1 one can find an upper bound
on the maximum eigenvalue of $8\boldsymbol{\boldsymbol{\boldsymbol{R}}}(\boldsymbol{\boldsymbol{\boldsymbol{z}}}^{k})$
and form $\boldsymbol{\boldsymbol{\boldsymbol{M}}}$ as: 

\begin{equation}
\boldsymbol{\boldsymbol{M}}=(m+s(P-1)^{1/2})\boldsymbol{\boldsymbol{I}}_{P}\label{eq:beig}
\end{equation}

where $m=\frac{8}{P}\text{Tr}(\boldsymbol{\boldsymbol{\boldsymbol{R}}}(\boldsymbol{\boldsymbol{\boldsymbol{z}}}^{k}))$,
$s^{2}=(\frac{64}{P}\text{Tr}(\boldsymbol{\boldsymbol{\boldsymbol{R}}}(\boldsymbol{\boldsymbol{\boldsymbol{z}}}^{k})^{2}))-m^{2}$.
Here on, we name the three approaches of obtaining $\boldsymbol{\boldsymbol{\boldsymbol{M}}}$
as TR (using TRace), EI (using EIgen value), BEI (using Bound on the
EIgen value). In the following we will explore another approach to
arrive at $\boldsymbol{\boldsymbol{\boldsymbol{M}}}$.

\textbf{Lemma-}2 {[}Lemma-3 and Lemma-4 \cite{21_MM_PSL}{]}: Let
$\boldsymbol{\boldsymbol{\boldsymbol{A}}}$ be an $P\times P$ Hermitian
Toeplitz matrix defined as follows

\[
\boldsymbol{\boldsymbol{\boldsymbol{A}}}=\begin{bmatrix}a(0) & a^{*}(1) & . & . & a^{*}(P-2) & a^{*}(P-1)\\
a(1) & a(0) & a^{*}(1) & . & . & a^{*}(P-2)\\
. & a(1) & a(0) & a^{*}(1) & . & .\\
. & . & a(1) & . & . & .\\
a(P-2) & . & . & . & . & a^{*}(1)\\
a(P-1) & a(P-2) & . & . & a(1) & a(0)
\end{bmatrix}
\]

and $\boldsymbol{\hat{\boldsymbol{E}}}^{H}$ be a $2P\times2P$ FFT
matrix with $\boldsymbol{\hat{\boldsymbol{E}}}(m,n)=e^{j\frac{2\Pi}{2P}mn},\hphantom{n}0\leq m,n\leq2P.$
Let $\boldsymbol{\boldsymbol{\boldsymbol{d}}}=[a_{0},a_{1},...,a_{P-1},0,a_{P-1}^{*},...,a_{1}^{*}]^{T}$
and $\boldsymbol{\boldsymbol{\boldsymbol{s}}}=\boldsymbol{\hat{\boldsymbol{E}}}^{H}\boldsymbol{\boldsymbol{\boldsymbol{d}}}$
be the discrete fourier transform of $\boldsymbol{\boldsymbol{\boldsymbol{d}}}$. 

(a) Then the maximum eigenvalue of the Hermitian Toeplitz matrix $\boldsymbol{\boldsymbol{\boldsymbol{A}}}$
can be bounded as

\begin{equation}
\lambda_{max}(\boldsymbol{\boldsymbol{\boldsymbol{A}}})\leq\frac{1}{2}\biggl(\underset{1\leq i\leq P}{\text{max}}\boldsymbol{\boldsymbol{\boldsymbol{s}}}_{2i}+\underset{1\leq i\leq P}{\text{max}}\boldsymbol{\boldsymbol{\boldsymbol{s}}}_{2i-1}\biggr)\label{eq:maxllamda}
\end{equation}

(b) The Hermitian Toeplitz matrix $\boldsymbol{\boldsymbol{\boldsymbol{A}}}$
can be decomposed as

\begin{equation}
\boldsymbol{\boldsymbol{\boldsymbol{A}}}=\frac{1}{2P}\boldsymbol{\boldsymbol{\hat{\boldsymbol{E}}}}_{:,1:P}\text{Diag}(\boldsymbol{\boldsymbol{\boldsymbol{d}}})\boldsymbol{\boldsymbol{\hat{\boldsymbol{E}}}}_{:,1:P}^{H}\label{eq:decompose}
\end{equation}

Proof: The proof can be find in \cite{21_MM_PSL} $\hphantom{nnnnnnnnnnnnnnnnnnnnnnnnnnnnnnnnnnnnnnnnnnnnnnnnn}$$\blacksquare$

Using Lemma-2, one can also find the bound on maximum eigenvalue of
a Hermitian Toeplitz matrix $8\boldsymbol{\boldsymbol{\boldsymbol{R}}}(\boldsymbol{\boldsymbol{\boldsymbol{z}}}^{k})$
using FFT and IFFT operations as:

\begin{equation}
\boldsymbol{\boldsymbol{M}}=4\biggl(\underset{1\leq i\leq P}{\text{max}}\boldsymbol{\boldsymbol{\boldsymbol{s}}}_{2i}+\underset{1\leq i\leq P}{\text{max}}\boldsymbol{\boldsymbol{\boldsymbol{s}}}_{2i-1}\biggr)\boldsymbol{\boldsymbol{I}}_{P}.\label{eq:lamfft}
\end{equation}

where $\boldsymbol{\boldsymbol{\boldsymbol{d}}}=[r(0),r(1),...,r(P-1),0,r(P-1)^{*},...,r(1)^{*}]^{T}$
and $\boldsymbol{\boldsymbol{\boldsymbol{s}}}=\boldsymbol{\hat{\boldsymbol{E}}}^{H}\boldsymbol{\boldsymbol{\boldsymbol{d}}}.$ 

We will name this approach as BEFFT (Bound on Eigenvalue using FFT). 

So, from (\ref{eq:st2}) we have the upper bound (majorization) function
of the original objective function $g(\boldsymbol{\boldsymbol{\boldsymbol{z}}})$
at any fixed point $\boldsymbol{\boldsymbol{\boldsymbol{z}}}^{k}$
as:

\begin{equation}
u(\boldsymbol{\boldsymbol{z}}|\boldsymbol{\boldsymbol{\boldsymbol{z}}}^{k})=\boldsymbol{\boldsymbol{\boldsymbol{z}}}^{H}(0.5\boldsymbol{\boldsymbol{\boldsymbol{M}}})\boldsymbol{\boldsymbol{\boldsymbol{z}}}+4\text{Re}((\boldsymbol{\boldsymbol{\boldsymbol{z}}}^{k})^{H}(\boldsymbol{\boldsymbol{\boldsymbol{R}}}(\boldsymbol{\boldsymbol{\boldsymbol{z}}}^{k})-0.25\boldsymbol{\boldsymbol{\boldsymbol{M}}})\boldsymbol{\boldsymbol{\boldsymbol{z}}})+(\boldsymbol{\boldsymbol{\boldsymbol{z}}}^{k})^{H}(0.5\boldsymbol{\boldsymbol{\boldsymbol{M}}}-3\boldsymbol{\boldsymbol{\boldsymbol{R}}}(\boldsymbol{\boldsymbol{\boldsymbol{z}}}^{k}))\boldsymbol{\boldsymbol{z}}^{k}\label{eq:majeq}
\end{equation}
As the $\boldsymbol{\boldsymbol{M}}$ (obtained by all four approaches
described above) is a constant times diagonal matrix and $\boldsymbol{\boldsymbol{\boldsymbol{z}}}^{H}\boldsymbol{\boldsymbol{\boldsymbol{z}}}$
being a constant, the first and last terms in the (\ref{eq:majeq})
are constants. So, after ignoring the constant terms, the surrogate
minimization problem can be rewritten as:

\begin{equation}
\begin{aligned} & \underset{\boldsymbol{\boldsymbol{\boldsymbol{z}}}}{\text{\text{minimize}}} &  & u(\boldsymbol{\boldsymbol{\boldsymbol{z}}}|\boldsymbol{\boldsymbol{\boldsymbol{z}}}^{k})=4\text{Re}((\boldsymbol{\boldsymbol{\boldsymbol{z}}}^{k})^{H}(\boldsymbol{\boldsymbol{\boldsymbol{R}}}(\boldsymbol{\boldsymbol{\boldsymbol{z}}}^{k})-0.25\boldsymbol{\boldsymbol{\boldsymbol{M}}})\boldsymbol{\boldsymbol{\boldsymbol{z}}})\\
 & \text{subject to} &  & |z_{n}|=1,\;n=1,...,P.
\end{aligned}
\label{eq:prob3}
\end{equation}

The problem in (\ref{eq:prob3}) can be rewritten more compactly as:

\begin{equation}
\begin{aligned} & \underset{\boldsymbol{\boldsymbol{\boldsymbol{z}}}}{\text{\text{minimize}}} &  & u(\boldsymbol{\boldsymbol{\boldsymbol{z}}}|\boldsymbol{\boldsymbol{\boldsymbol{z}}}^{k})=||\boldsymbol{\boldsymbol{\boldsymbol{z}}}-\boldsymbol{\tilde{\boldsymbol{\boldsymbol{a}}}}||_{2}^{2}\\
 & \text{subject to} &  & |z_{n}|=1,\;n=1,...,P,
\end{aligned}
\label{eq:prob4}
\end{equation}

where $\boldsymbol{\tilde{\boldsymbol{\boldsymbol{a}}}}=-(\boldsymbol{\boldsymbol{\boldsymbol{R}}}(\boldsymbol{\boldsymbol{\boldsymbol{z}}}^{k})-0.25\boldsymbol{\boldsymbol{\boldsymbol{M}}})\boldsymbol{\boldsymbol{\boldsymbol{z}}}^{k}$,
which involves computing Hermitian Toeplitz matrix-vector multiplication.
By using decomposition of a Toeplitz matrix (\ref{eq:decompose}),
one can easily implement it using FFT and IFFT operations.

The problem in (\ref{eq:prob4}) has a closed-form solution of 
\begin{equation}
\boldsymbol{\boldsymbol{\boldsymbol{z}}}^{k+1}=\frac{\boldsymbol{\tilde{\boldsymbol{\boldsymbol{a}}}}}{||\boldsymbol{\tilde{\boldsymbol{\boldsymbol{a}}}}||_{2}}.\label{eq:38}
\end{equation}

The pseudocode of the proposed algorithm-FISL is given below

\begin{algorithm}[h]
\textbf{Require}: sequence length $\text{\textquoteleft}P\text{\textquoteright}$

1:set $k=0$, initialize $\boldsymbol{z}^{0}$

2:\textbf{ repeat}

3:$\hphantom{nn}$compute $\boldsymbol{\boldsymbol{\boldsymbol{R}}}(\boldsymbol{\boldsymbol{\boldsymbol{z}}}^{k})$
using (\ref{eq:midMat})

4:$\hphantom{nn}$compute $\boldsymbol{\boldsymbol{\boldsymbol{M}}}$
using (\ref{eq:lamfft})

5:$\hphantom{nn}$$\boldsymbol{\tilde{\boldsymbol{\boldsymbol{a}}}}=-(\boldsymbol{\boldsymbol{\boldsymbol{R}}}(\boldsymbol{\boldsymbol{\boldsymbol{z}}}^{k})-0.25\boldsymbol{\boldsymbol{\boldsymbol{M}}})\boldsymbol{\boldsymbol{\boldsymbol{z}}}^{k}$

6:$\hphantom{nn}$$\boldsymbol{\boldsymbol{\boldsymbol{z}}}^{k+1}=\frac{\boldsymbol{\tilde{\boldsymbol{\boldsymbol{a}}}}}{||\boldsymbol{\tilde{\boldsymbol{\boldsymbol{a}}}}||_{2}}$

7:$\hphantom{nn}$ $k\longleftarrow k+1$

8:\textbf{ until }convergence

\caption{:FISL -Faster ISL minimization}
\end{algorithm}

\subsection*{B. Convergence analysis}

The proposed algorithm (FISL) is derived based on the MM technique.
The working principle of the MM technique is explained in the section
I-B. From (\ref{eq:MM-1}), we have

\[
g(\boldsymbol{\boldsymbol{\boldsymbol{z}}}^{k+1})\leq u(\boldsymbol{\boldsymbol{\boldsymbol{z}}}^{k+1}|\boldsymbol{\boldsymbol{\boldsymbol{z}}}^{k})\leq u(\boldsymbol{\boldsymbol{\boldsymbol{z}}}^{k}|\boldsymbol{\boldsymbol{z}}^{k})=g(\boldsymbol{\boldsymbol{\boldsymbol{z}}}^{k})
\]

So, MM technique is ensuring that the cost function value evaluated
at every point $\{\boldsymbol{\boldsymbol{\boldsymbol{z}}}^{k}\}$
generated by the FISL algorithm will be monotonically decreasing and
by the nature of the cost function of the problem in (\ref{eq:mainprob}),
one can observe that it is always bounded below by zero. So, the sequence
of cost function values is guaranteed to converge to a finite value.

Now, we will discuss the convergence of points $\{\boldsymbol{\boldsymbol{\boldsymbol{z}}}^{k}\}$
generated by the FISL algorithm to a stationary point. So, starting
with the definition of a stationary point.

\textbf{Proposition} 1: Let $f:\boldsymbol{\boldsymbol{\boldsymbol{R}}^{n}}\rightarrow\boldsymbol{\boldsymbol{\boldsymbol{R}}}$
be any smooth function and let $\boldsymbol{\boldsymbol{\boldsymbol{x}}}^{*}$
be a local minimum of $f$ over a subset $\boldsymbol{\boldsymbol{\boldsymbol{\chi}}}$
of $\boldsymbol{\boldsymbol{R}}^{n}$ \cite{bertsekas_convex}. Then

\begin{equation}
\nabla f(\boldsymbol{\boldsymbol{\boldsymbol{x}}}^{*})\boldsymbol{\boldsymbol{\boldsymbol{y}}}\geq0,\:\forall\boldsymbol{\boldsymbol{\boldsymbol{y}}}\in T_{\boldsymbol{\boldsymbol{\boldsymbol{\chi}}}}(\boldsymbol{\boldsymbol{\boldsymbol{x}}}^{*})\label{eq:sp}
\end{equation}

where $T_{\boldsymbol{\boldsymbol{\boldsymbol{\chi}}}}(\boldsymbol{\boldsymbol{\boldsymbol{x}}}^{*})$
denotes the tangent cone of $\boldsymbol{\boldsymbol{\boldsymbol{\chi}}}$
at $\boldsymbol{\boldsymbol{\boldsymbol{x}}}^{*}$. Such any point
$\boldsymbol{\boldsymbol{\boldsymbol{x}}}^{*}$, which satisfies (\ref{eq:sp})
is called as a stationary point.

Now, the convergence property of the FISL algorithm is explained as
follows.

\textbf{Theorem} 2: Let $\left\{ \boldsymbol{\boldsymbol{\boldsymbol{z}}}^{k}\right\} $
be the sequence of points generated by the FISL algorithm. Then every
point $\left\{ \boldsymbol{\boldsymbol{\boldsymbol{z}}}^{k}\right\} $
is a stationary point of the problem in (\ref{eq:mainprob}).

Proof: Assume that there exists a converging subsequence $\boldsymbol{\boldsymbol{\boldsymbol{z}}}^{l_{j}}\rightarrow\boldsymbol{\boldsymbol{\boldsymbol{z}}}^{*}$,
then from the theory of MM technique, we have

\[
u(\boldsymbol{\boldsymbol{\boldsymbol{z}}^{(l_{j+1})}}|\boldsymbol{\boldsymbol{z}}^{(l_{j+1})})=g(\boldsymbol{\boldsymbol{z}}^{(l_{j+1})})\leq g(\boldsymbol{\boldsymbol{z}}^{(l_{j}+1)})\leq u(\boldsymbol{\boldsymbol{\boldsymbol{z}}^{(l_{j}+1)}}|\boldsymbol{\boldsymbol{z}}^{(l_{j})})\leq u(\boldsymbol{\boldsymbol{\boldsymbol{z}}}|\boldsymbol{\boldsymbol{z}}^{(l_{j})})
\]

\[
u(\boldsymbol{\boldsymbol{\boldsymbol{z}}^{(l_{j+1})}}|\boldsymbol{\boldsymbol{z}}^{(l_{j+1})})\leq u(\boldsymbol{\boldsymbol{\boldsymbol{z}}}|\boldsymbol{\boldsymbol{z}}^{(l_{j})})
\]

Letting $j\rightarrow+\infty$, we obtain

\begin{equation}
u(\boldsymbol{\boldsymbol{\boldsymbol{z}}^{\infty}}|\boldsymbol{\boldsymbol{z}}^{\infty})\leq u(\boldsymbol{\boldsymbol{\boldsymbol{z}}}|\boldsymbol{\boldsymbol{z}}^{\infty})\label{eq:sp2}
\end{equation}

Replacing $\boldsymbol{\boldsymbol{z}}^{\infty}\text{ with }\boldsymbol{\boldsymbol{\boldsymbol{z}}}^{*}$,
we have

\begin{equation}
u(\boldsymbol{\boldsymbol{\boldsymbol{z}}}^{*}|\boldsymbol{\boldsymbol{\boldsymbol{z}}}^{*})\leq u(\boldsymbol{\boldsymbol{\boldsymbol{z}}}|\boldsymbol{\boldsymbol{\boldsymbol{z}}}^{*})\label{eq:sp3}
\end{equation}

So, (\ref{eq:sp3}) conveys that $\boldsymbol{\boldsymbol{\boldsymbol{z}}}^{*}$
is a stationary point and also a global minimizer of $u(.)$ i.e.,

\begin{equation}
\nabla u(\boldsymbol{\boldsymbol{\boldsymbol{z}}}^{*})\boldsymbol{\boldsymbol{\boldsymbol{d}}}\geq0,\:\forall\boldsymbol{\boldsymbol{\boldsymbol{d}}}\in T_{\boldsymbol{\boldsymbol{\boldsymbol{Z}}}}(\boldsymbol{\boldsymbol{\boldsymbol{z}}}^{*})\label{eq:spg}
\end{equation}

From the majorization step, we know that the first-order behavior
of majorized function $u(\boldsymbol{\boldsymbol{\boldsymbol{z}}}|\boldsymbol{\boldsymbol{\boldsymbol{z}}}^{k})$
is equal to the original cost function $g(\boldsymbol{\boldsymbol{\boldsymbol{z}}})$.
So, we can show

\begin{equation}
u(\boldsymbol{\boldsymbol{\boldsymbol{z}}}^{*}|\boldsymbol{\boldsymbol{\boldsymbol{z}}}^{*})\leq u(\boldsymbol{\boldsymbol{\boldsymbol{z}}}|\boldsymbol{\boldsymbol{\boldsymbol{z}}}^{*})\Leftrightarrow g(\boldsymbol{\boldsymbol{\boldsymbol{z}}}^{*})\leq g(\boldsymbol{\boldsymbol{\boldsymbol{z}}})\label{eq:spg2}
\end{equation}

and it leads to

\begin{equation}
\nabla g(\boldsymbol{\boldsymbol{\boldsymbol{z}}}^{*})\boldsymbol{\boldsymbol{\boldsymbol{y}}}\geq0,\forall\boldsymbol{\boldsymbol{\boldsymbol{y}}}\in T_{\boldsymbol{\boldsymbol{\boldsymbol{Z}}}}(\boldsymbol{\boldsymbol{\boldsymbol{z}}}^{*})\label{eq:spg3}
\end{equation}

So, the set of points generated by the FISL algorithm are stationary
points and $\boldsymbol{\boldsymbol{\boldsymbol{z}}}^{*}$ is the
minimizer of $g(\boldsymbol{\boldsymbol{\boldsymbol{z}}})$. This
concludes the proof.$\hphantom{nnnnnnnnnnnnnnnnnnnnnnnnnnnnnnnnnnnnnnnnnnnnnnnnnnnnnnnnnnnnnn}$$\blacksquare$

\subsection*{C. Computational $\&$ Space Complexity}

The per iteration computational complexity of the proposed algorithm
(FISL) is dominated in forming a Hermitian Toeplitz matrix $\boldsymbol{\boldsymbol{\boldsymbol{R}}}(\boldsymbol{\boldsymbol{\boldsymbol{z}}}^{k})$,
Diagonal matrix $\boldsymbol{\boldsymbol{\boldsymbol{M}}}$ and Hermitian
Toeplitz matrix-vector multiplication to form $\boldsymbol{\tilde{\boldsymbol{\boldsymbol{a}}}}$.
But by using the Lemma-2, we replaced all of them using FFT and IFFT
operations, and to implement our algorithm we require only 3-FFT and
2-IFFT operations and the computational complexity would be $\mathcal{O}(P\,log\,P)$.
In each iteration of our algorithm, the space complexity is dominated
by the three different vectors of sizes $P\times1$, $(2P-1)\times1$
, $2P\times1$, respectively and the space complexity would be $\mathcal{O}(P).$
The computational \& space complexity of state-of-the-art algorithms
are given as: CAN-$\mathcal{O}(P\,log\,P)$, $\mathcal{O}(P)$, MISL-$\mathcal{O}(P\,log\,P)$,
$\mathcal{O}(P)$, ISL-NEW-$\mathcal{O}(P\,log\,P)$, $\mathcal{O}(P^{2})$,
ADMM-$\mathcal{O}(P^{3})$, $\mathcal{O}(P^{2})$, CPM-$\mathcal{O}(K\,log\,P)$,
$\mathcal{O}(P^{2})$ where $K\in\text{number of iterations in the bisection method}$.
Hence, our proposed algorithm has either same or better computational
\& space complexity than the state-of-the-art algorithms.

\section*{\centerline{III.NUMERICAL EXPERIMENTS}}

In this section, we will show the potential of our proposed algorithm
Faster ISL minimization (FISL) through some numerical simulations.
All simulations were performed in MATLAB on a laptop with a 2.50GHz
i7 processor. Experiments has been conducted for different sequence
lengths of $P=100,225,400,625,900,1225$ using different initializations
like Golomb sequence \cite{15_polyphaseseq_Zhang}, Frank sequence
\cite{13_Polyphasecode_frank}, random sequence, and to stop all the
algorithms, we use the following convergence criterion:

\begin{equation}
\Biggr|\frac{(\text{ISL}(k+1)-\text{ISL}(k))}{\text{max}(1,\text{ISL}(k))}\Biggr|\leq10^{-5},\label{eq:40}
\end{equation}

where $\text{ISL}(k)$ is the ISL metric value at $k^{th}$ iteration.
In the case of random initialization, for every length each experiment
is repeated for 30 Monte Carlo trials and for each trial different
random initial sequence is used i.e., $\boldsymbol{\boldsymbol{\boldsymbol{z}}}^{0}$
is chosen as $\bigl\{ e^{j2\pi\theta_{i}}\bigr\}_{i=1}^{P}$, where
$\bigl\{\theta_{i}\bigr\}$ are drawn randomly from the uniform distribution
$\left[0,1\right]$.

In each experiment, the performance of the designed sequence such
as ISL metric value, auto-correlation side-lobe levels and algorithm
performance in terms of convergence speed to reach the stationary
point is observed and compared with the state-of-the-art algorithms
like CAN \cite{16_CAN}, MISL \cite{17_MISL}, ISL-NEW \cite{20_fast_alg_waveform_LI},
ADMM approach \cite{19_ADMM}, and CPM \cite{CPM_kehrodi}. First,
we will show the comparison of different approaches to construct the
matrix $\boldsymbol{\boldsymbol{\boldsymbol{M}}}$, which plays a
major role in the majorization step of our algorithm.

\begin{figure}[tph]
\subfloat[$P=100$]{\includegraphics[scale=0.55]{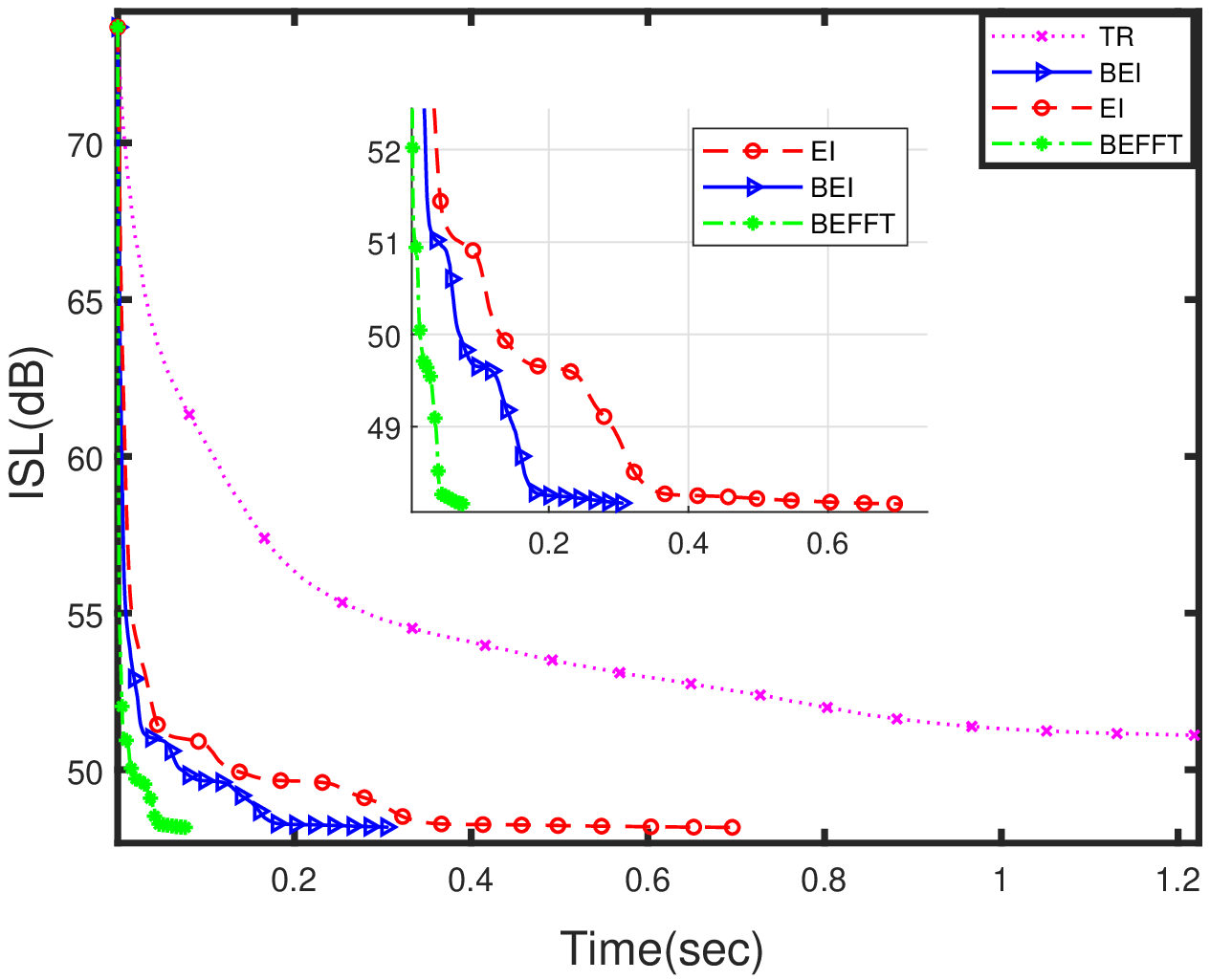}}\subfloat[$P=1225$]{\includegraphics[scale=0.55]{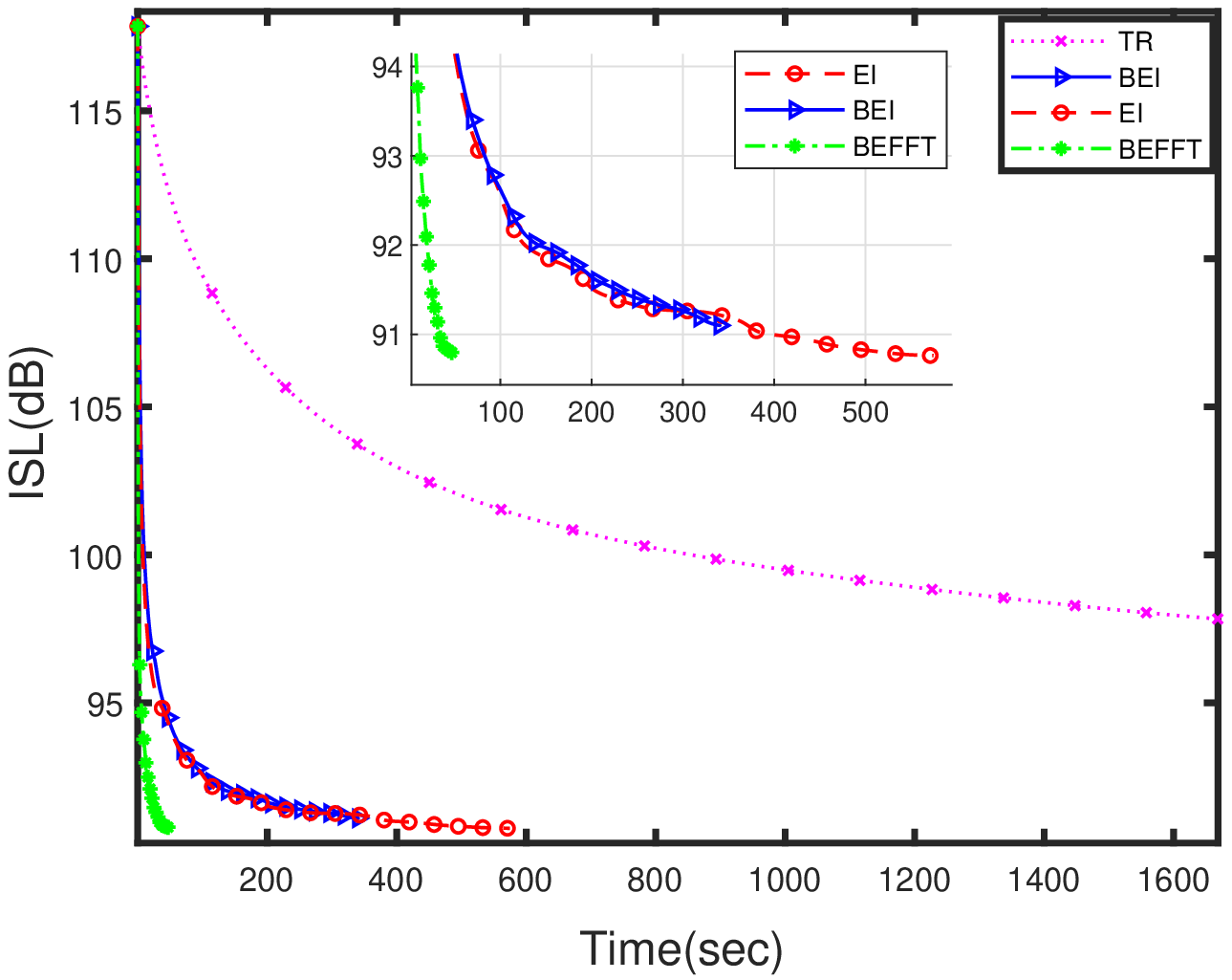}}

\subfloat[$P=100$]{\includegraphics[scale=0.55]{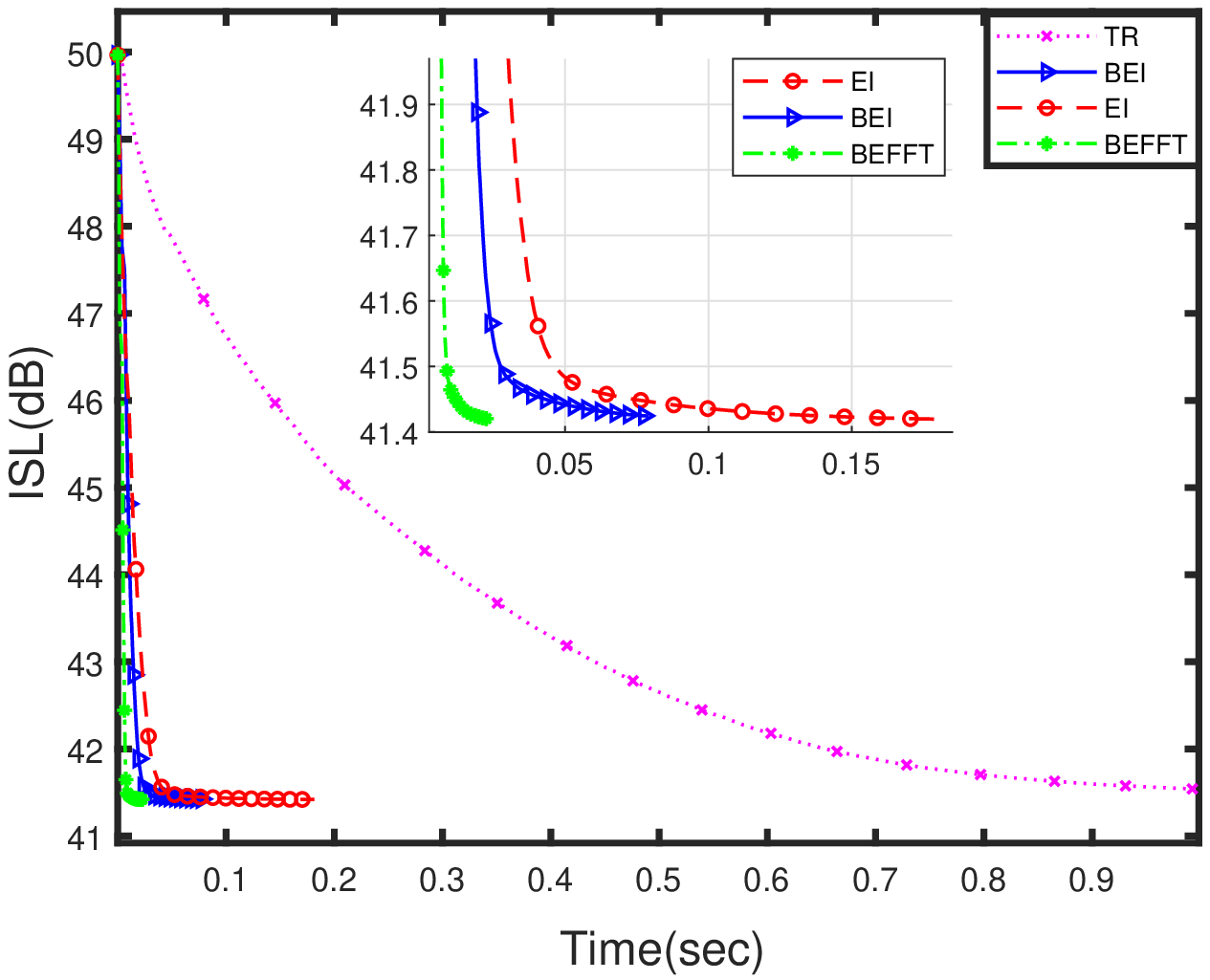}}\subfloat[$P=1225$]{\includegraphics[scale=0.55]{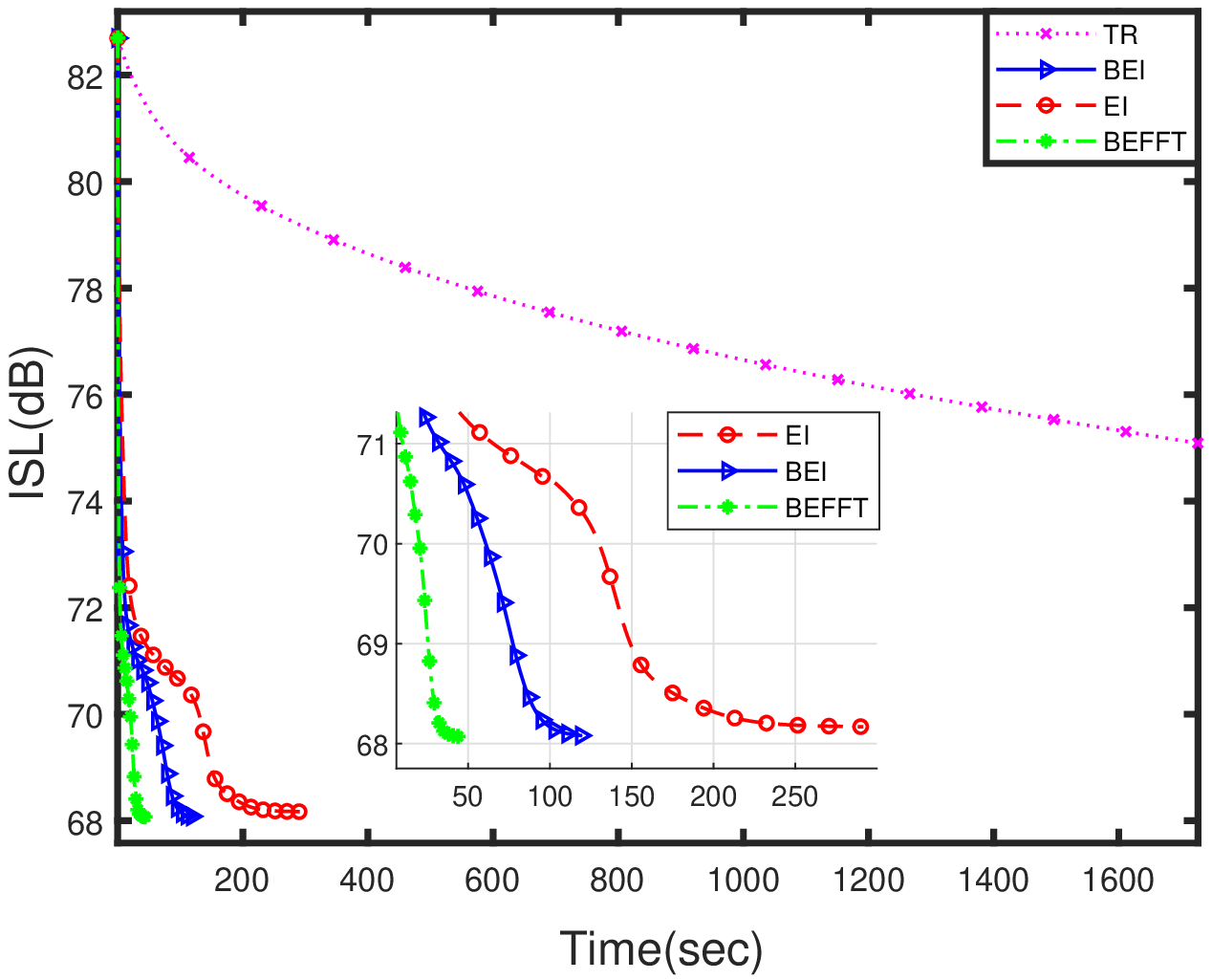}}

\subfloat[$P=100$]{\includegraphics[scale=0.55]{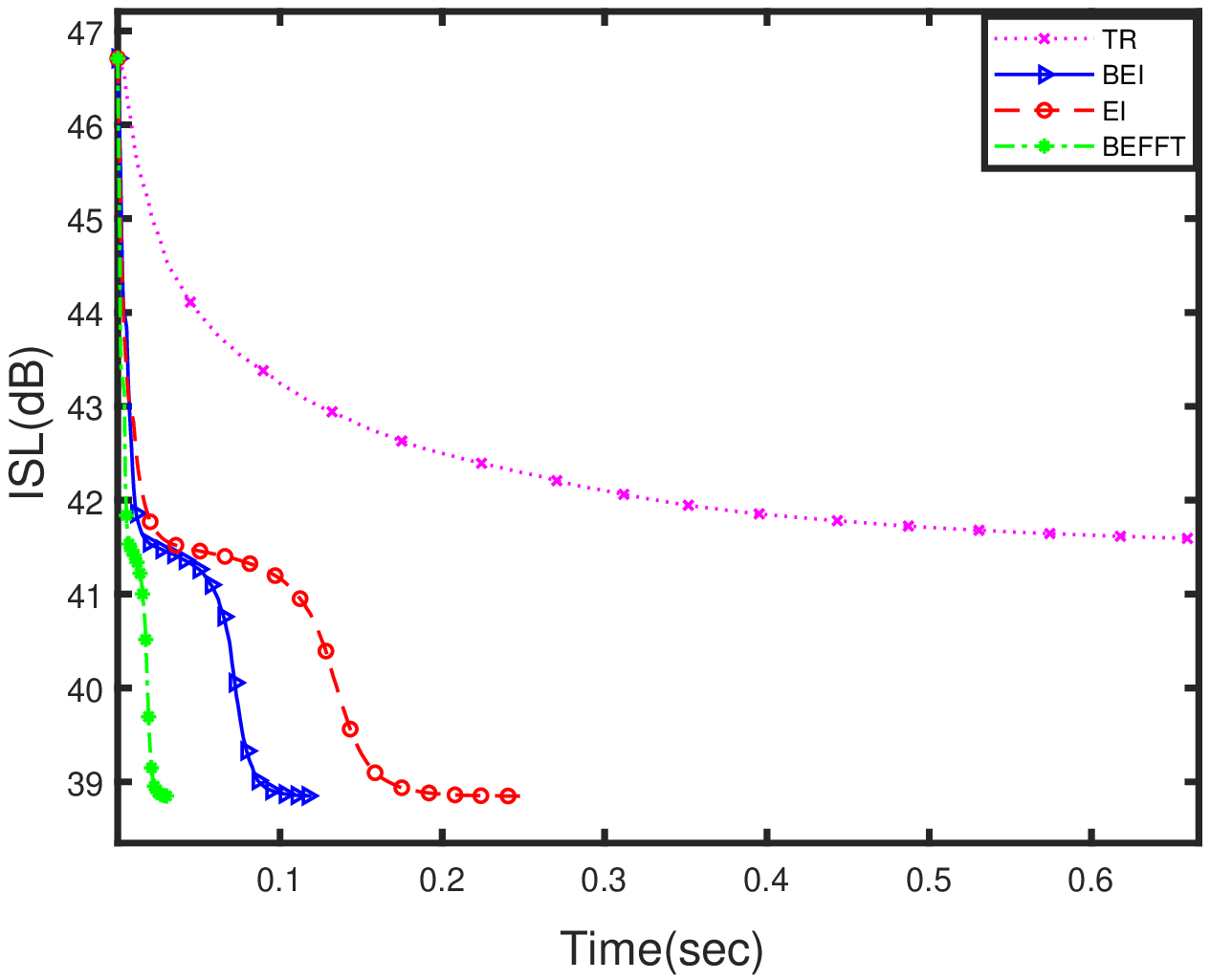}}\subfloat[$P=1225$]{\includegraphics[scale=0.55]{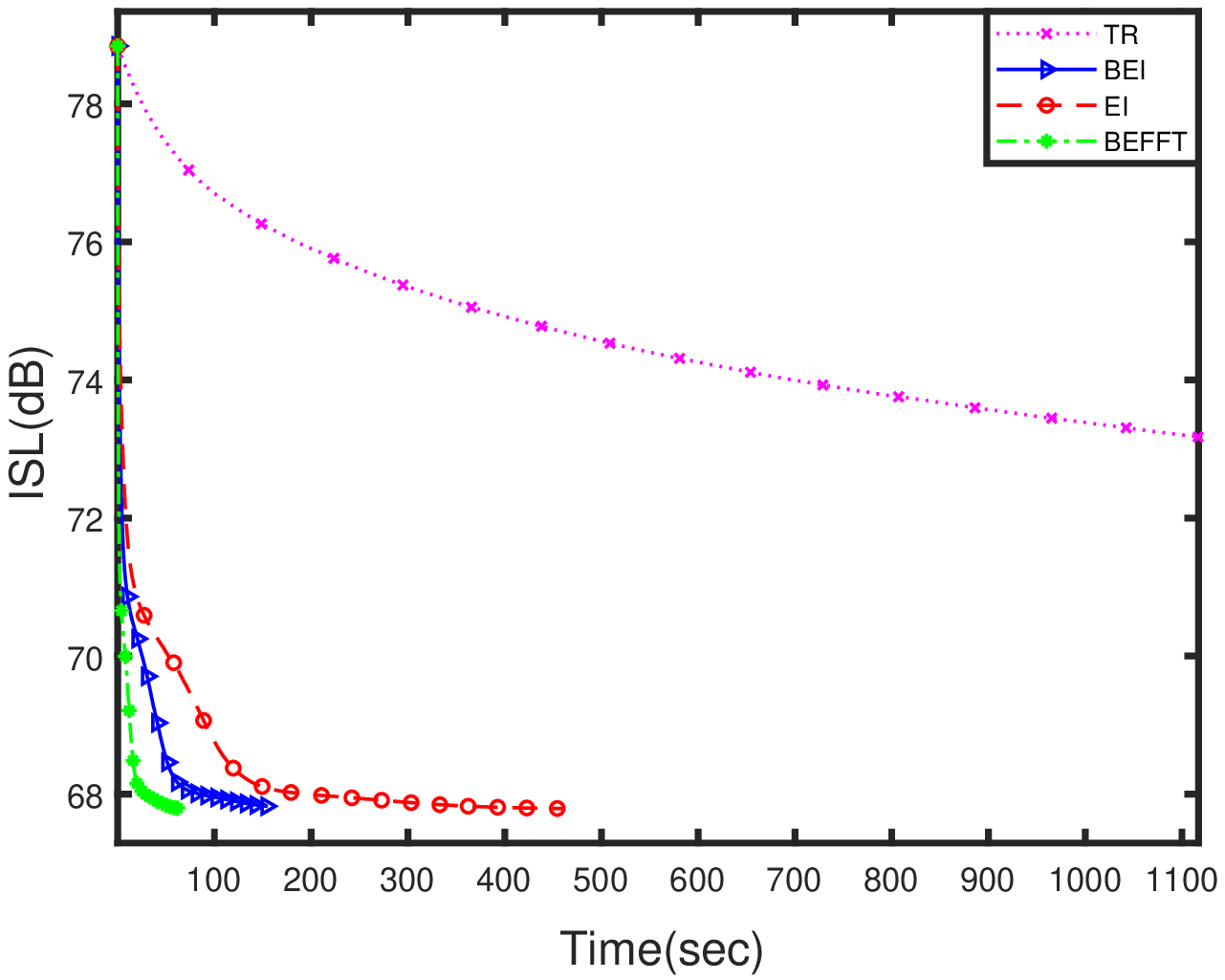}}

\caption{ISL with respect to time for sequence lengths $P=100,1225$. (a) and
(b) are for initialization via Random sequence. (c) and (d) are for
initialization via Golomb sequence. (e) and (f) are for initialization
via Frank sequence.}
\end{figure}

\begin{figure}[tph]
\subfloat[$P=100$]{\includegraphics[scale=0.55]{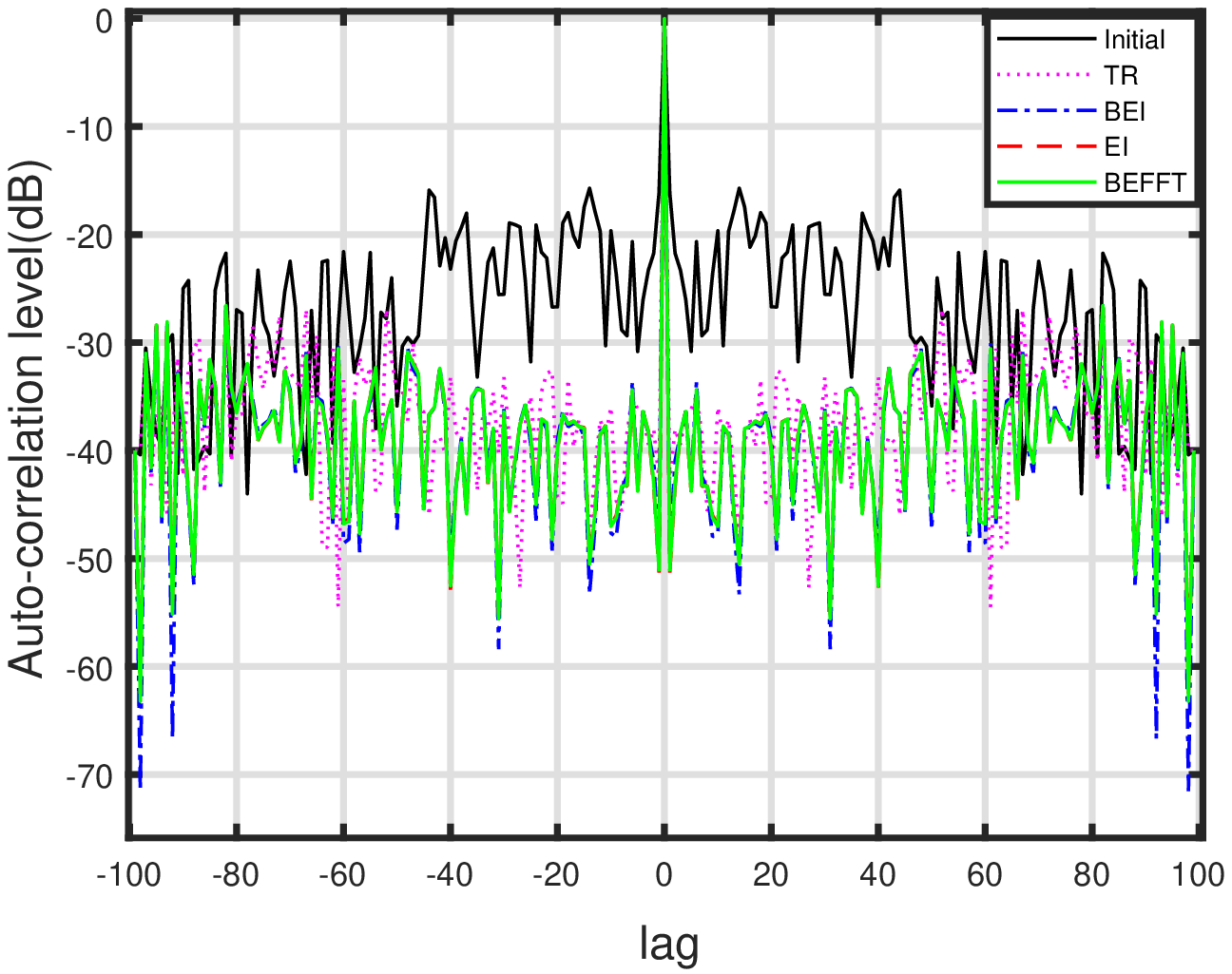}}\subfloat[$P=1225$]{\includegraphics[scale=0.55]{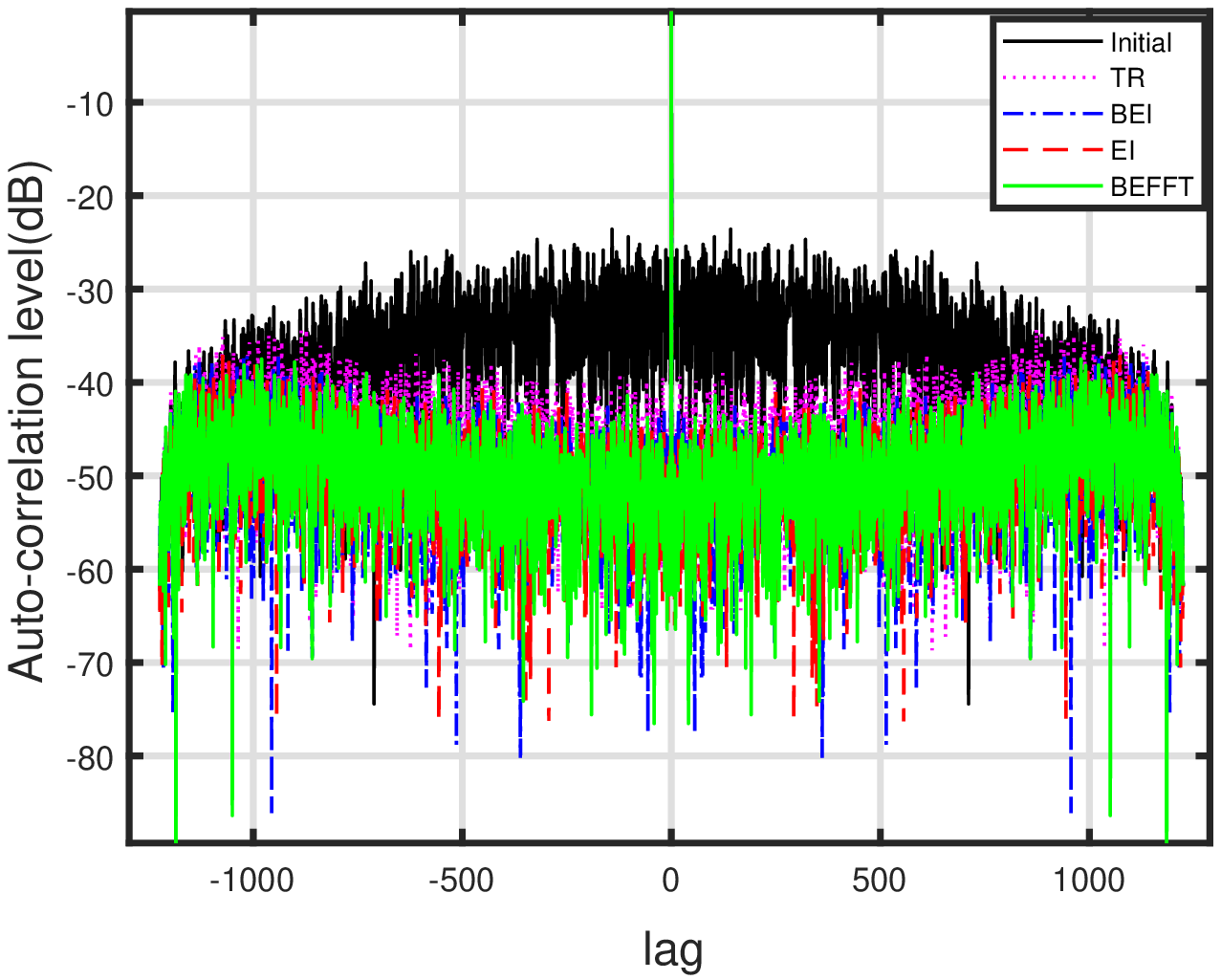}}

\subfloat[$P=100$]{\includegraphics[scale=0.55]{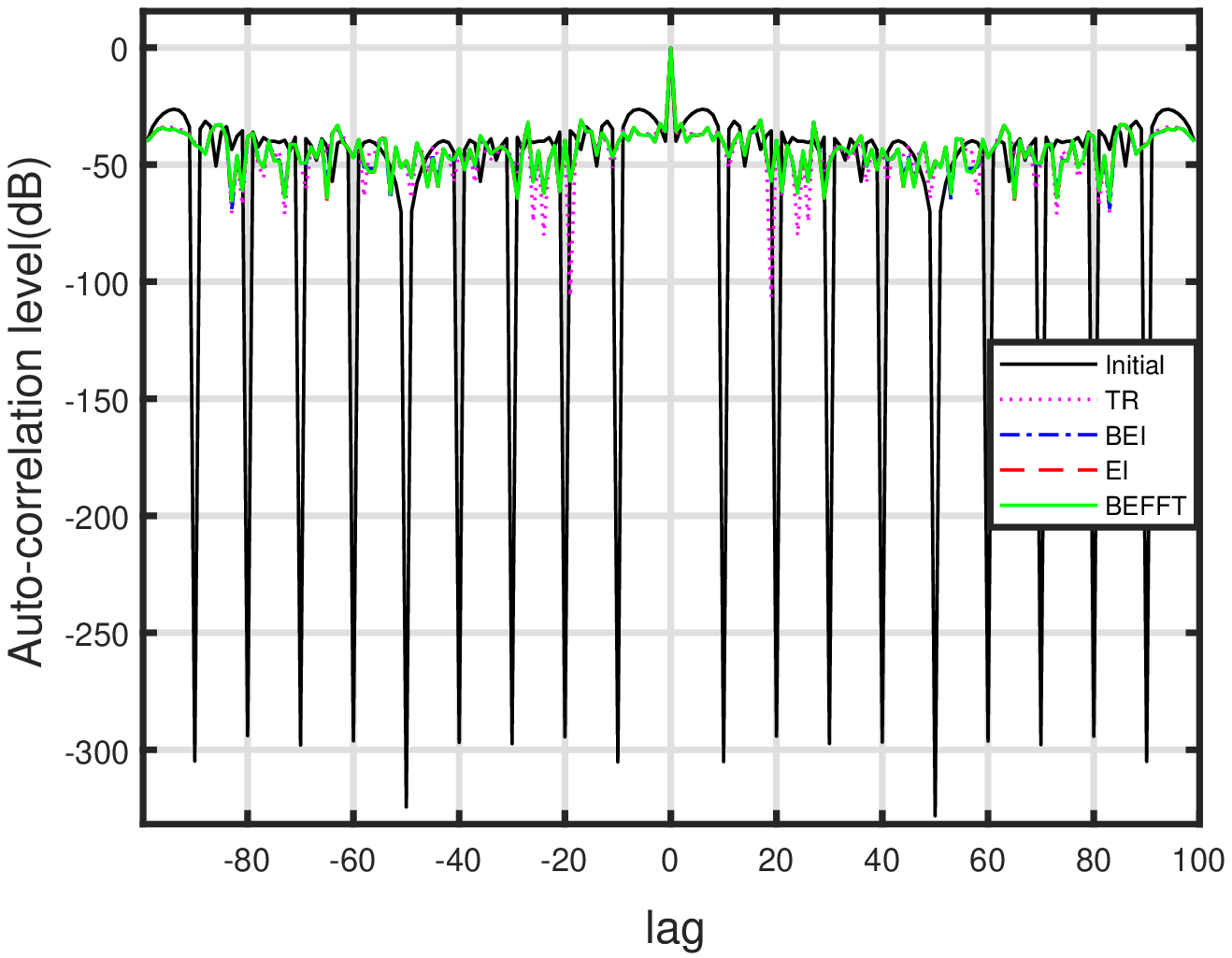}}\subfloat[$P=1225$]{\includegraphics[scale=0.55]{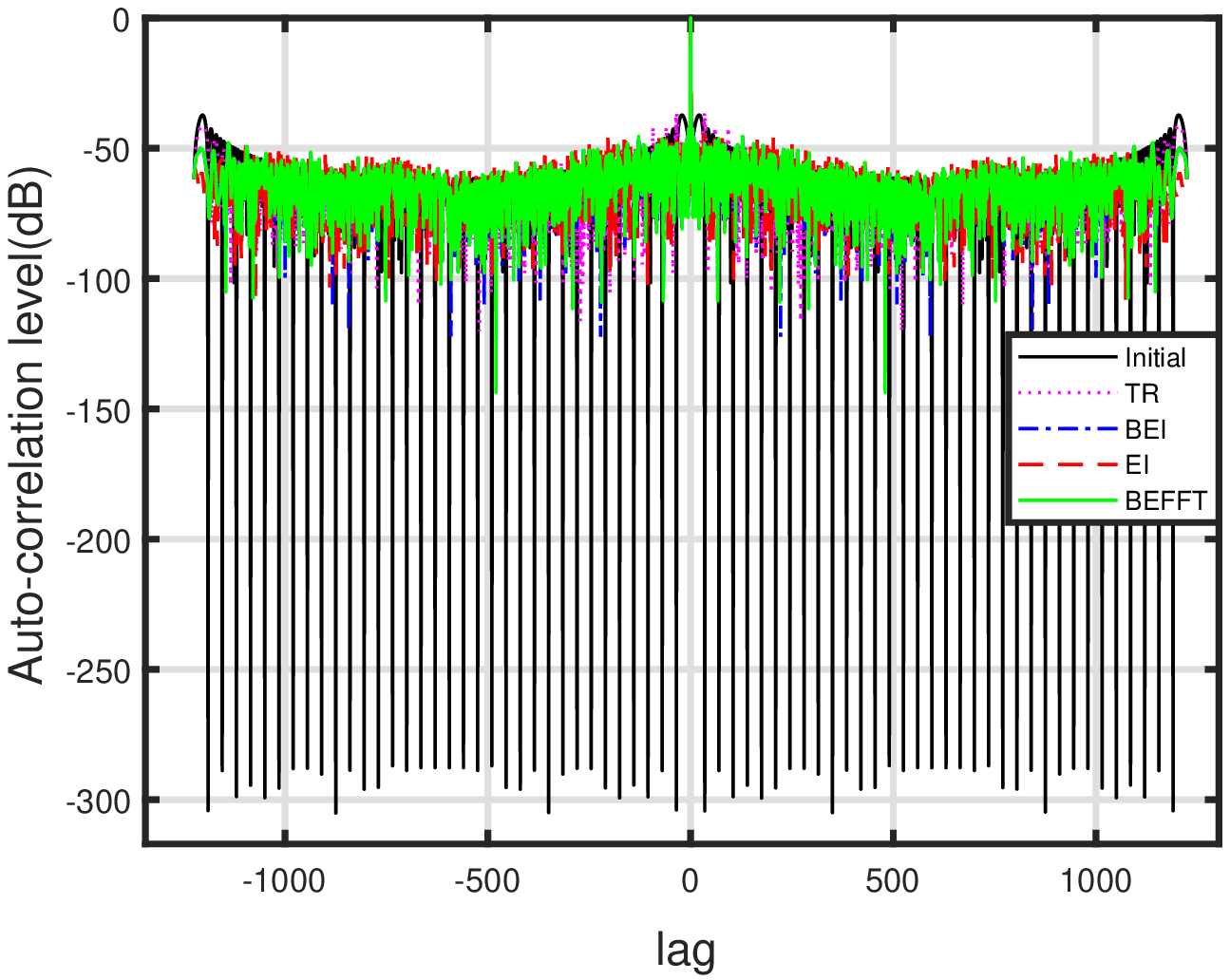}}

\subfloat[$P=100$]{\includegraphics[scale=0.55]{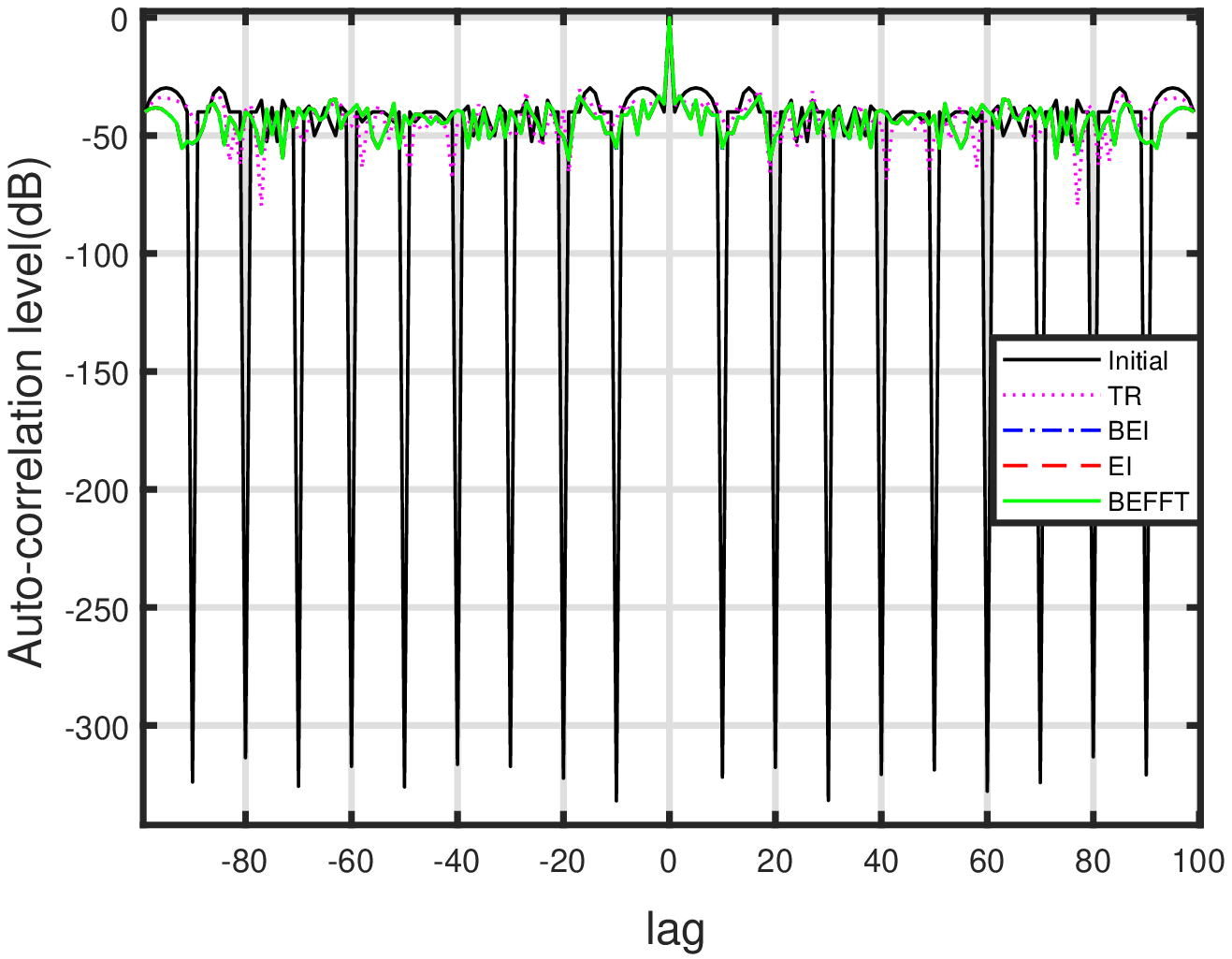}}\subfloat[$P=1225$]{\includegraphics[scale=0.55]{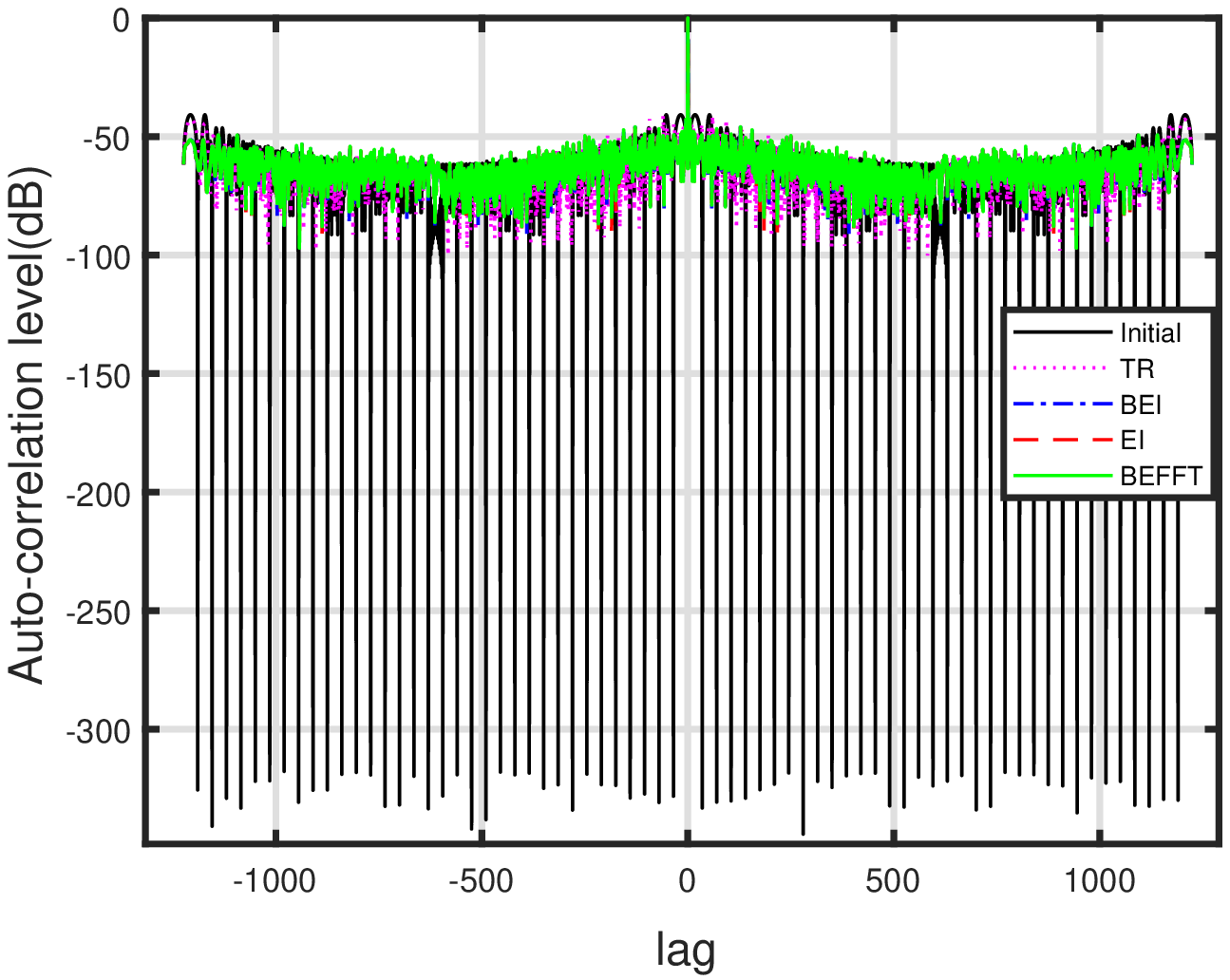}}

\caption{Auto-correlation value with respect to lag for sequence lengths $P=100,1225$.
(a) and (b) are for initialization via Random sequence. (c) and (d)
are for initialization via Golomb sequence. (e) and (f) are for initialization
via Frank sequence.}
\end{figure}

Figures. 1, 2 shows the normal and zoomed version (where ever it is
necessary) plots of ISL value vs time, auto-correlation value vs lag
for different sequence lengths $P=100,1225$ using three different
initializations, respectively. From the simulation plots, we observe
that, for all the initializations, all the approaches to construct
a matrix $\boldsymbol{\boldsymbol{\boldsymbol{M}}}$ will give the
same auto-correlation function but their convergence times to reach
minimum ISL value are different. In plots, we have shown results of
four different ways to construct $\boldsymbol{\boldsymbol{\boldsymbol{M}}}$
namely TR (i.e, by using an approach of TRace of a matrix), EI (i.e,
by using maximum EIgenvalue), BEI (i.e, by using Bound on the maximum
EIgen value), and BEFFT (i.e, by using Bound on the maximum Eigenvalue
using FFT operations). Among the four approaches, irrespective of
length and initialization, the BEFFT approach seems to have faster
convergence. From figure-1(b), one can observe that the BEFFT approach
is faster than TR, EI, BEI approaches by $38,13,8$ times respectively.
So, in the following, we have used only the BEFFT approach in the
update steps of our FISL algorithm.

Now, we will compare the performance of our FISL algorithm with the
state-of-the-art algorithms in terms of the ISL metric value, convergence
time, and auto-correlation side-lobe levels. For better comparison,
for each experiment, all the algorithms are initialized with the same
sequence and stopped using the same convergence criterion.

\begin{figure}[tph]
\subfloat[$P=100$]{\includegraphics[scale=0.55]{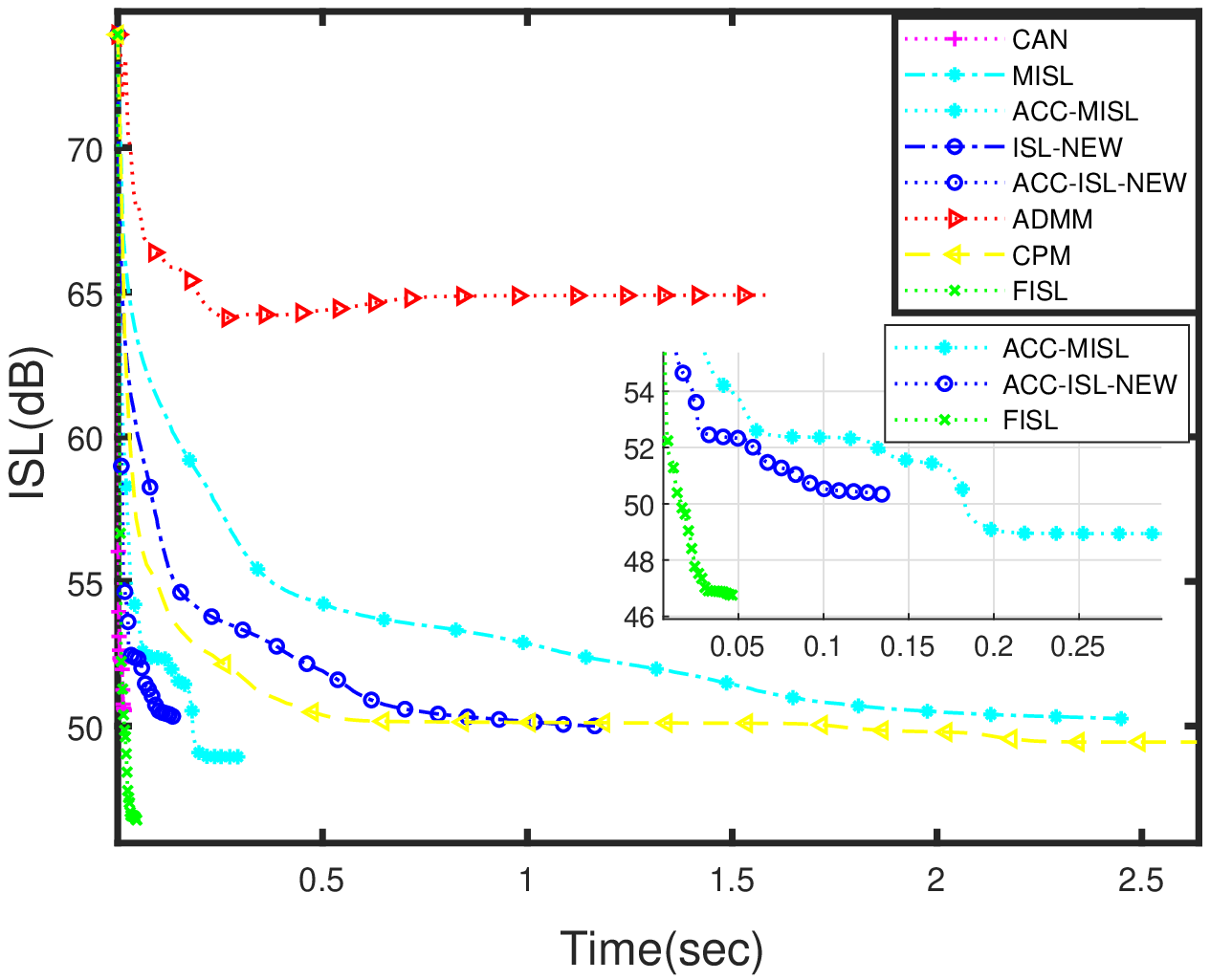}}\subfloat[$P=1225$]{\includegraphics[scale=0.55]{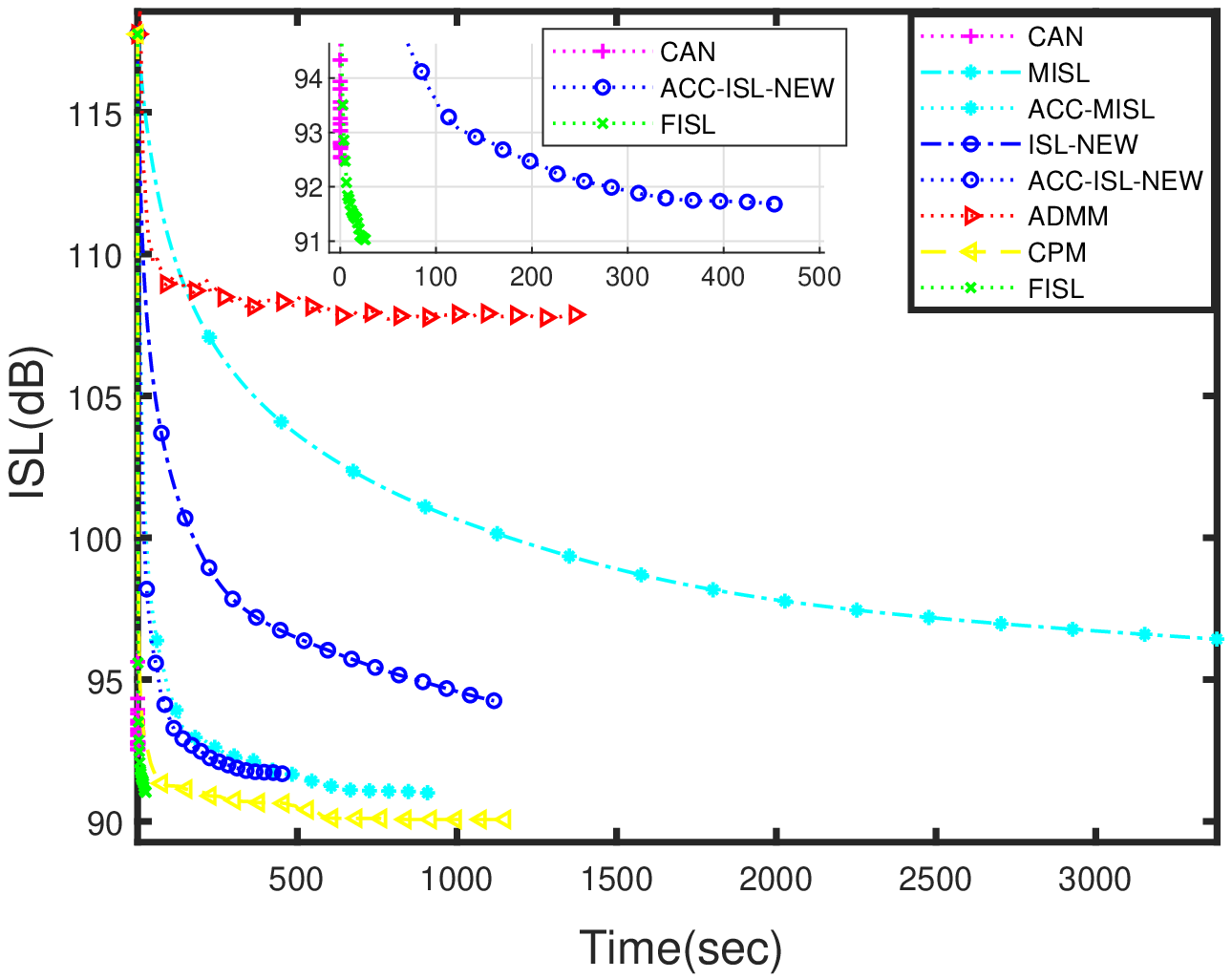}}

\subfloat[$P=100$]{\includegraphics[scale=0.55]{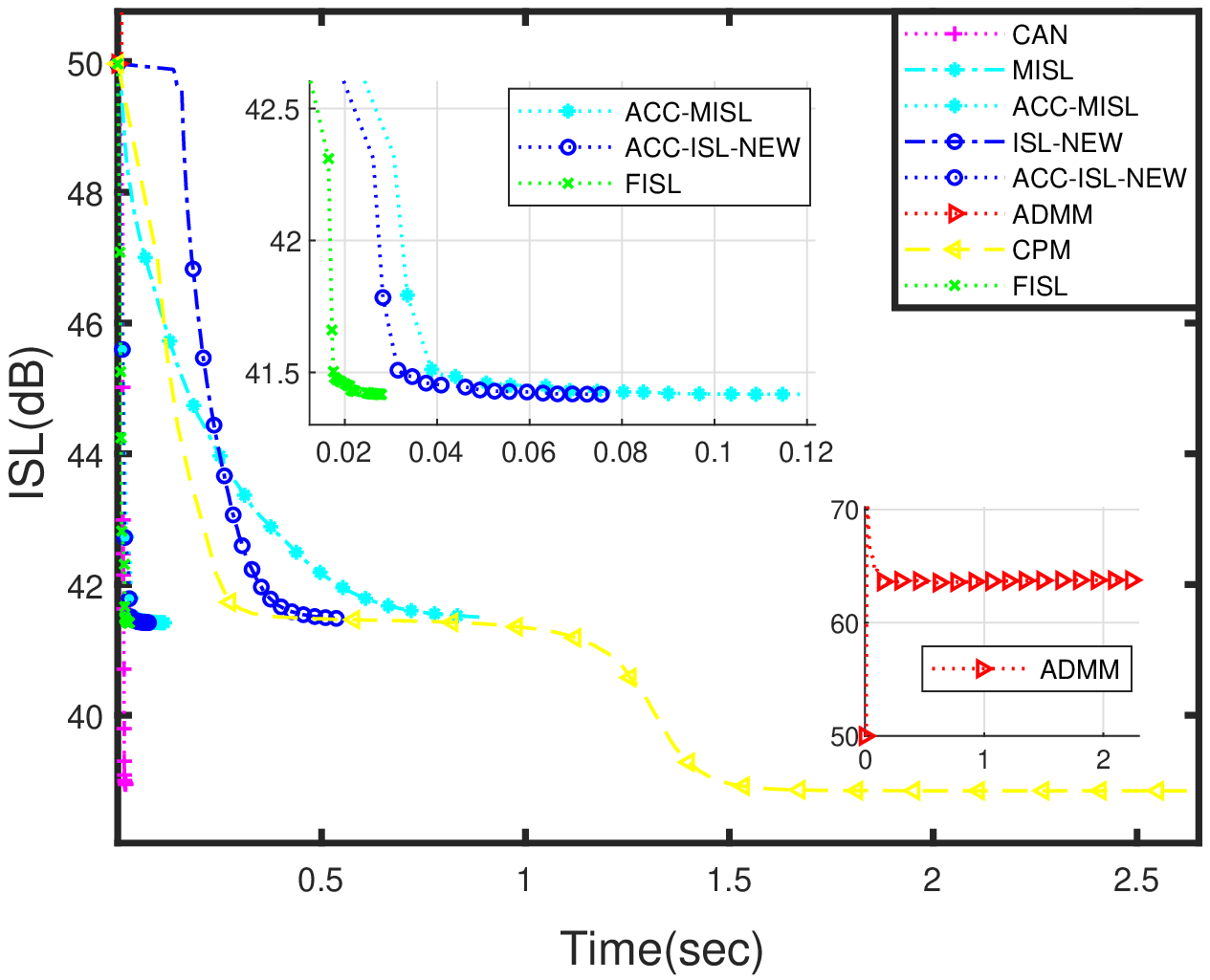}}\subfloat[$P=1225$]{\includegraphics[scale=0.55]{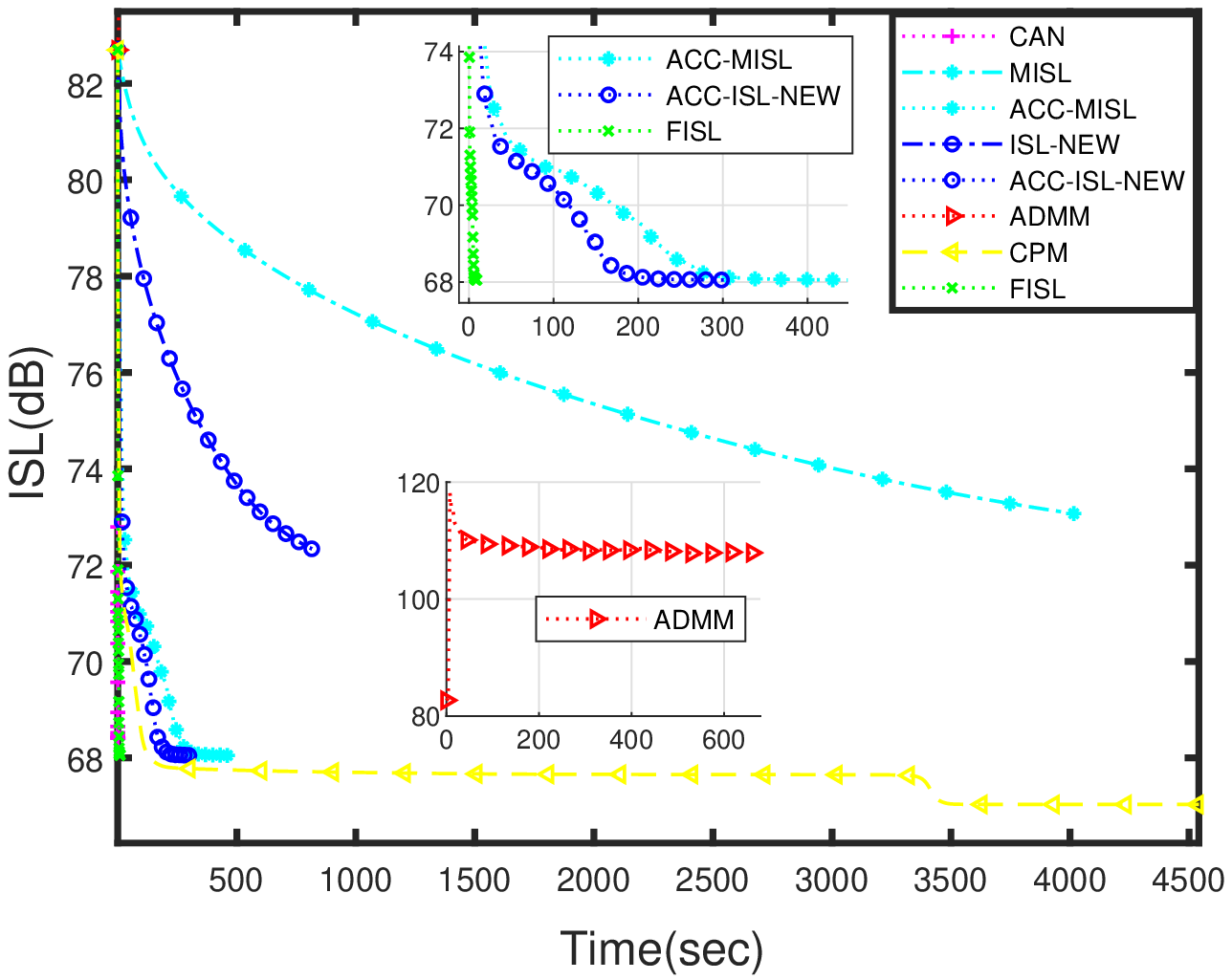}}

\subfloat[$P=100$]{\includegraphics[scale=0.55]{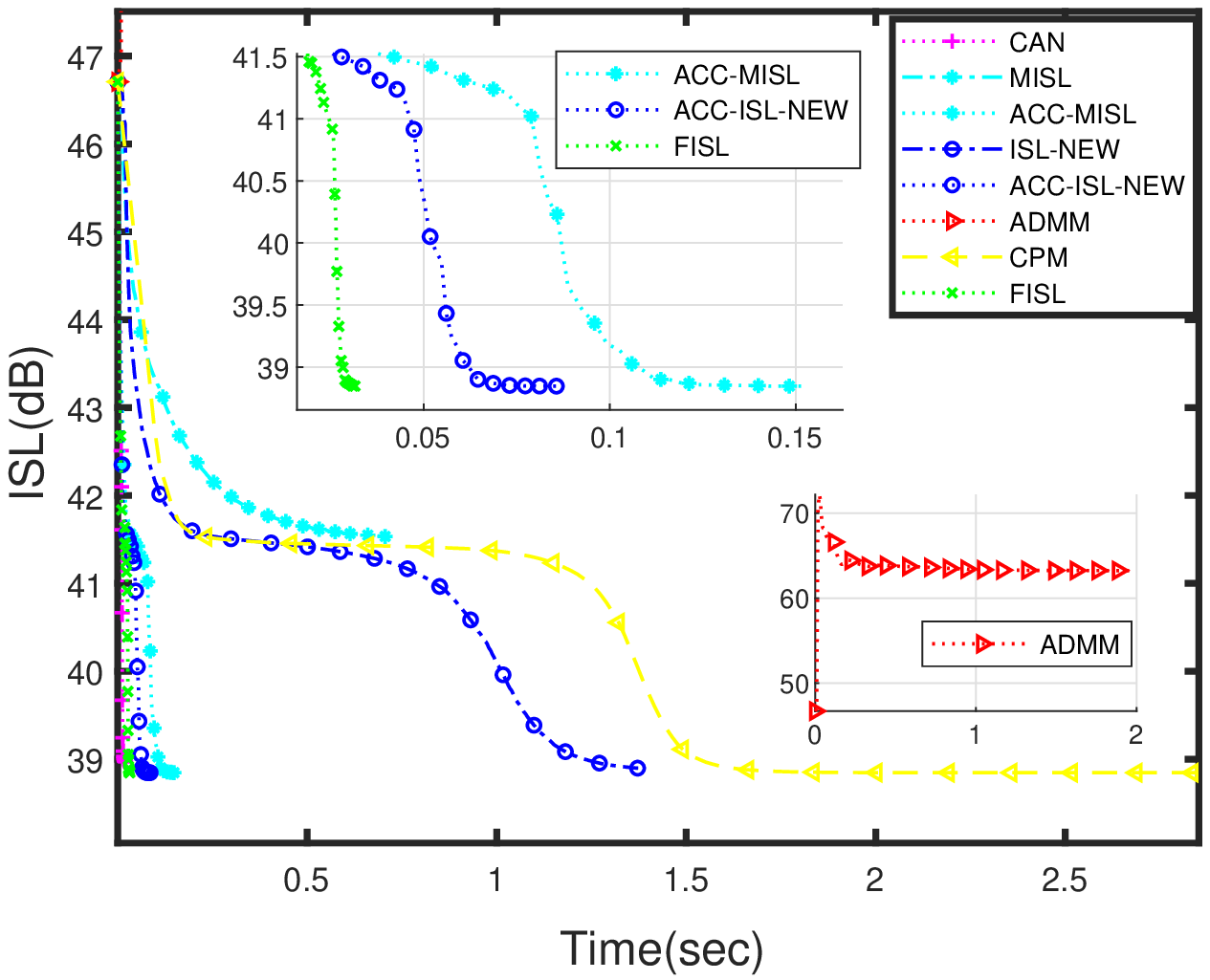}}\subfloat[$P=1225$]{\includegraphics[scale=0.55]{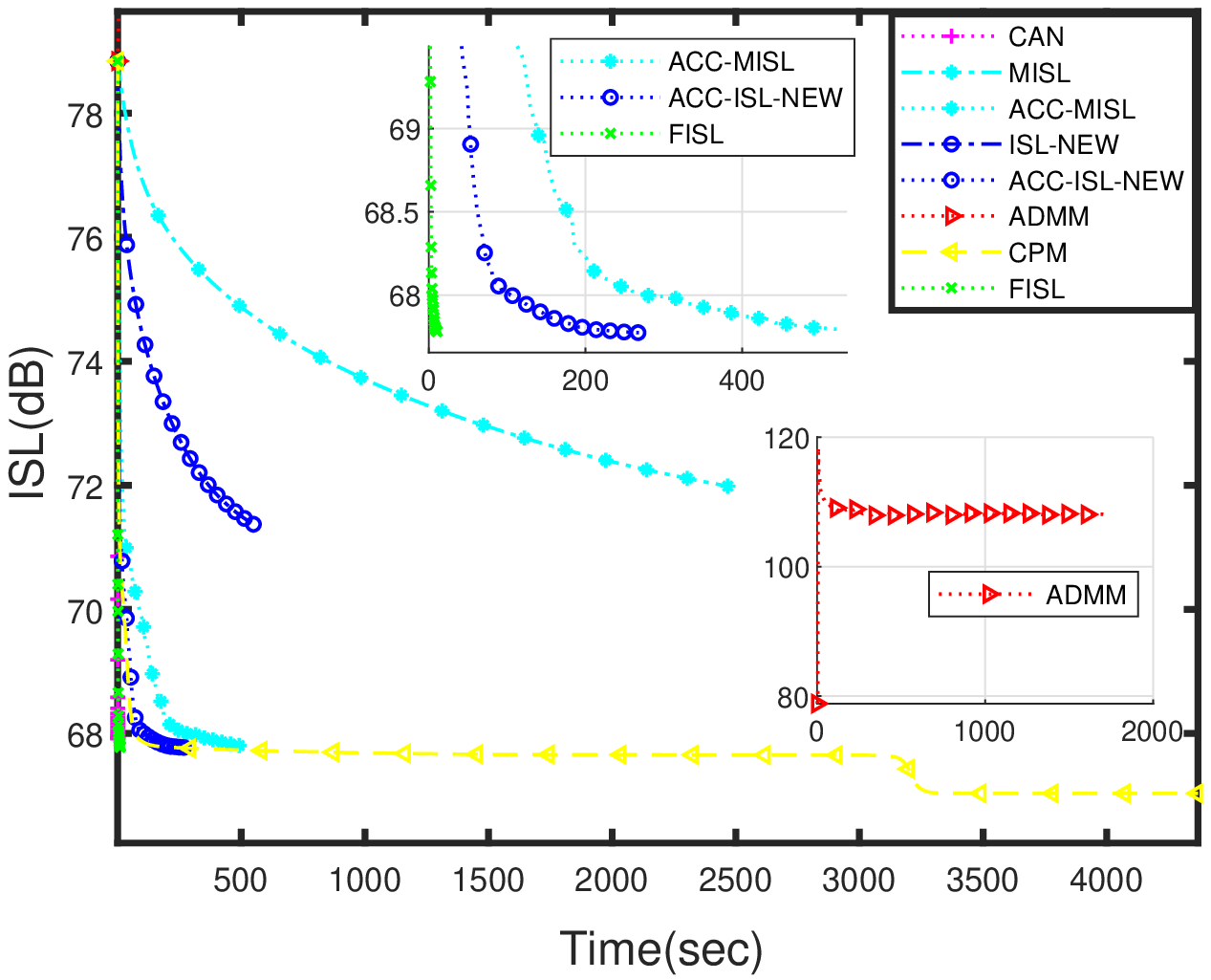}}

\caption{ISL with respect to time for a sequence length $P=100,1225$. (a)
and (b) are for initialization via Random sequence. (c) and (d) are
for initialization via Golomb sequence. (e) and (f) are for initialization
via Frank sequence.}
\end{figure}

\begin{figure}[tph]
\subfloat[$P=100$]{\includegraphics[scale=0.55]{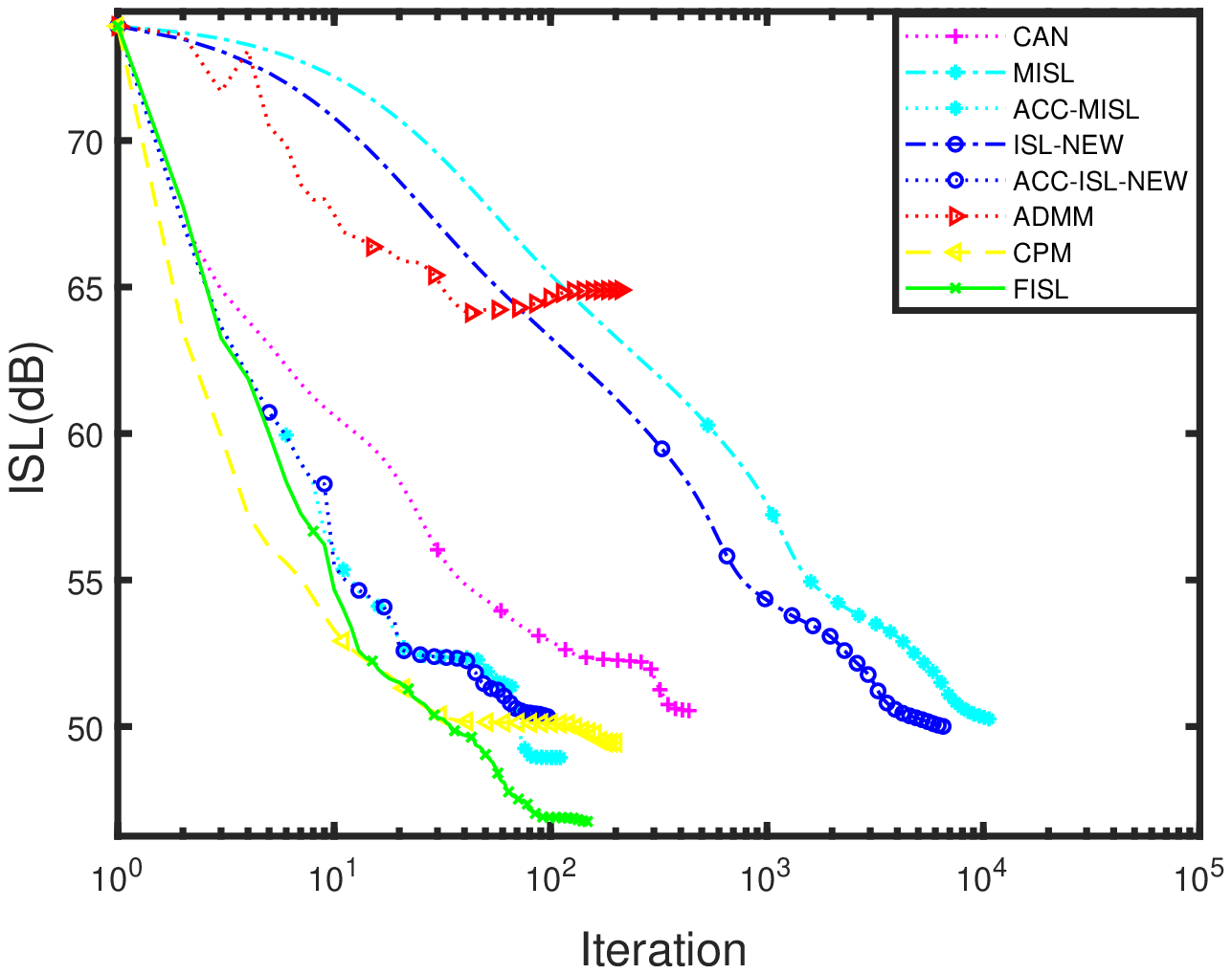}}\subfloat[$P=1225$]{\includegraphics[scale=0.55]{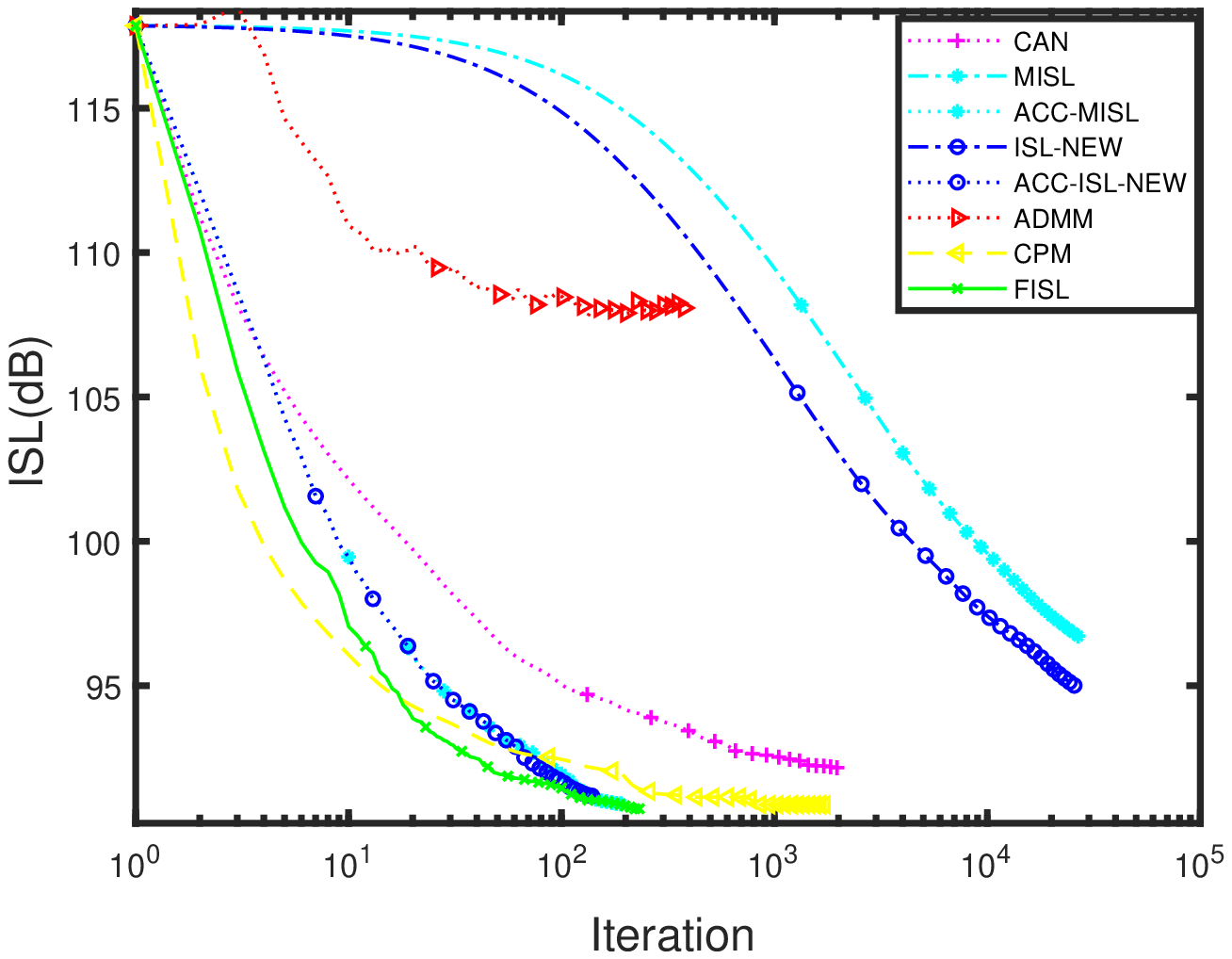}}

\subfloat[$P=100$]{\includegraphics[scale=0.55]{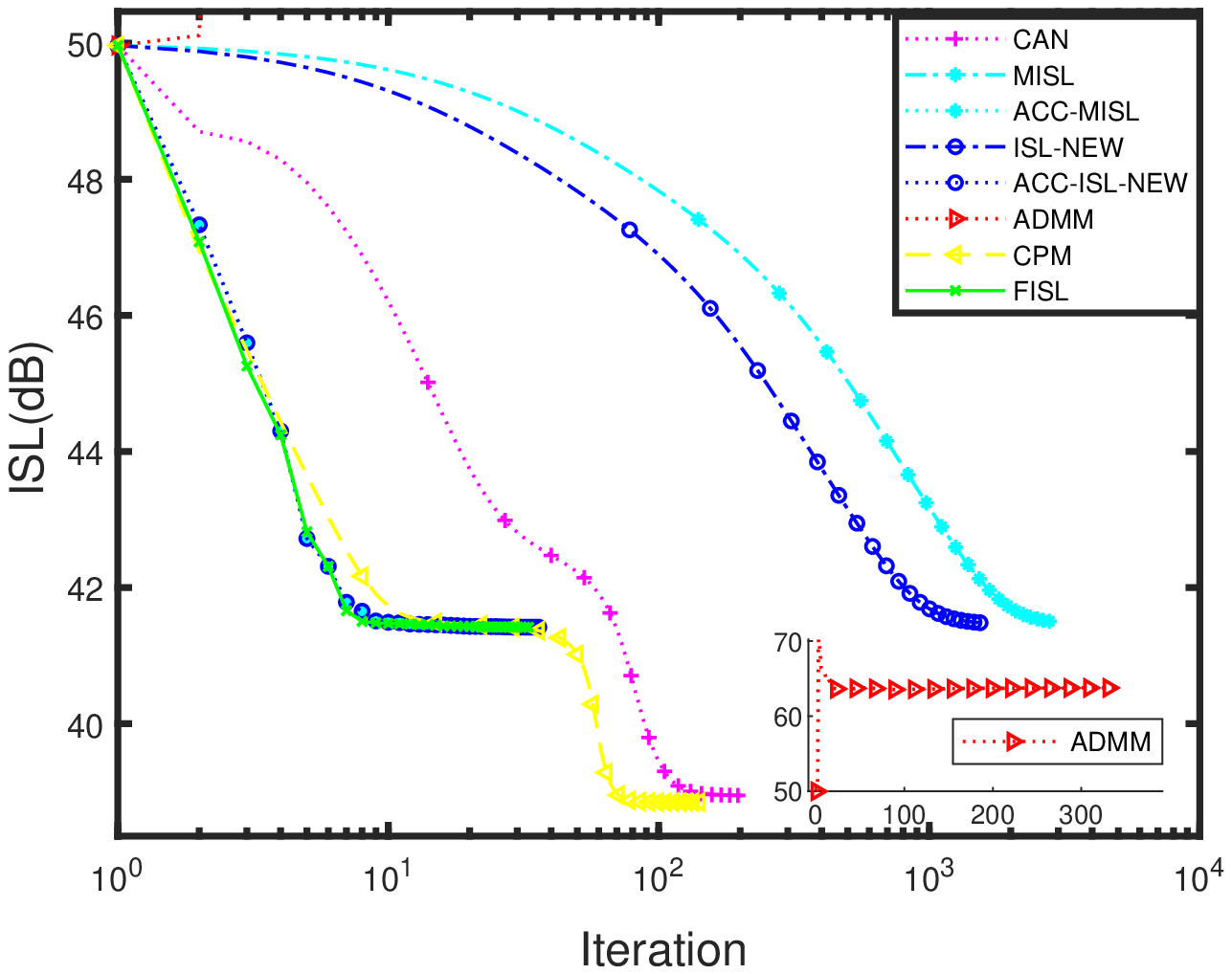}}\subfloat[$P=1225$]{\includegraphics[scale=0.55]{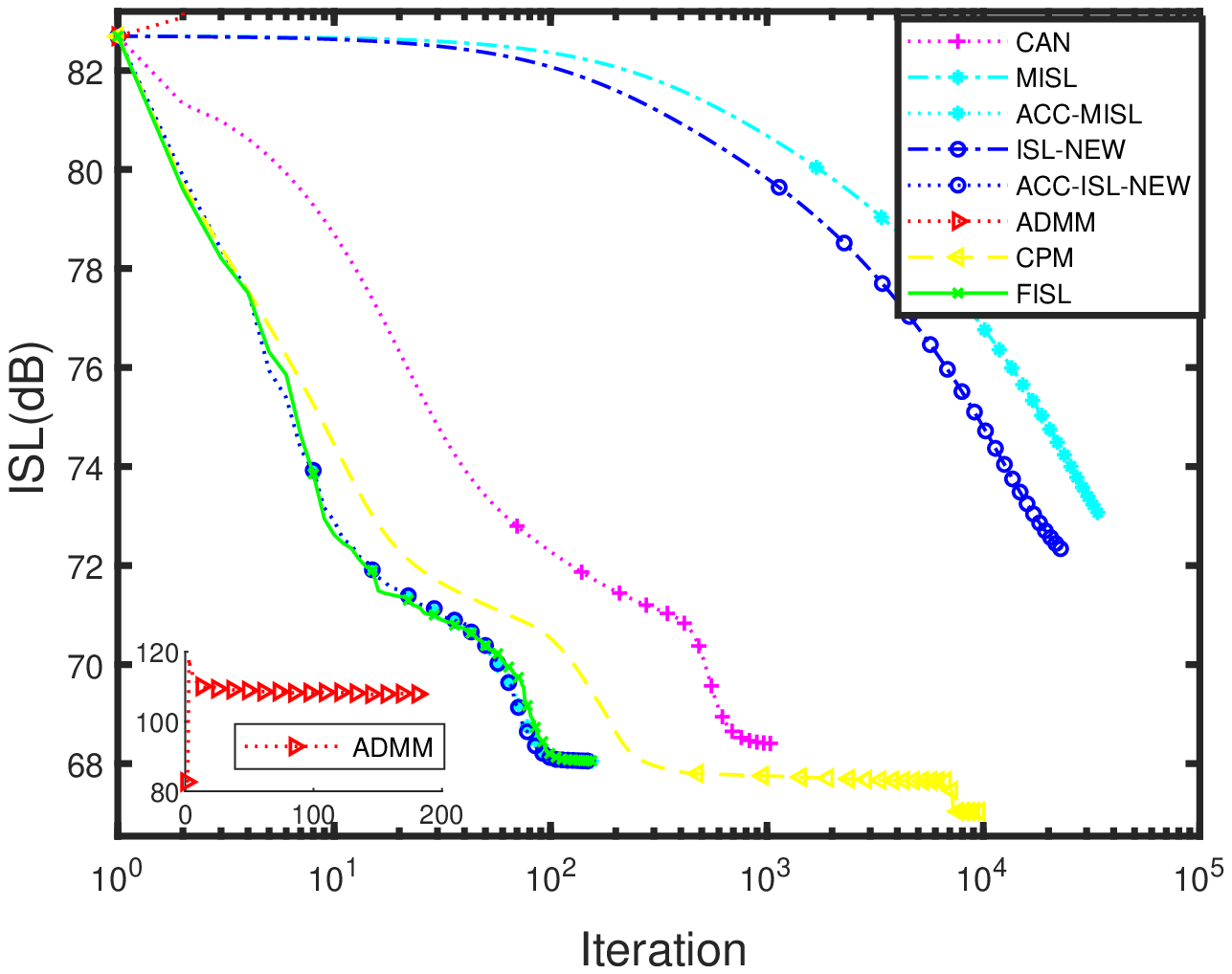}}

\subfloat[$P=100$]{\includegraphics[scale=0.55]{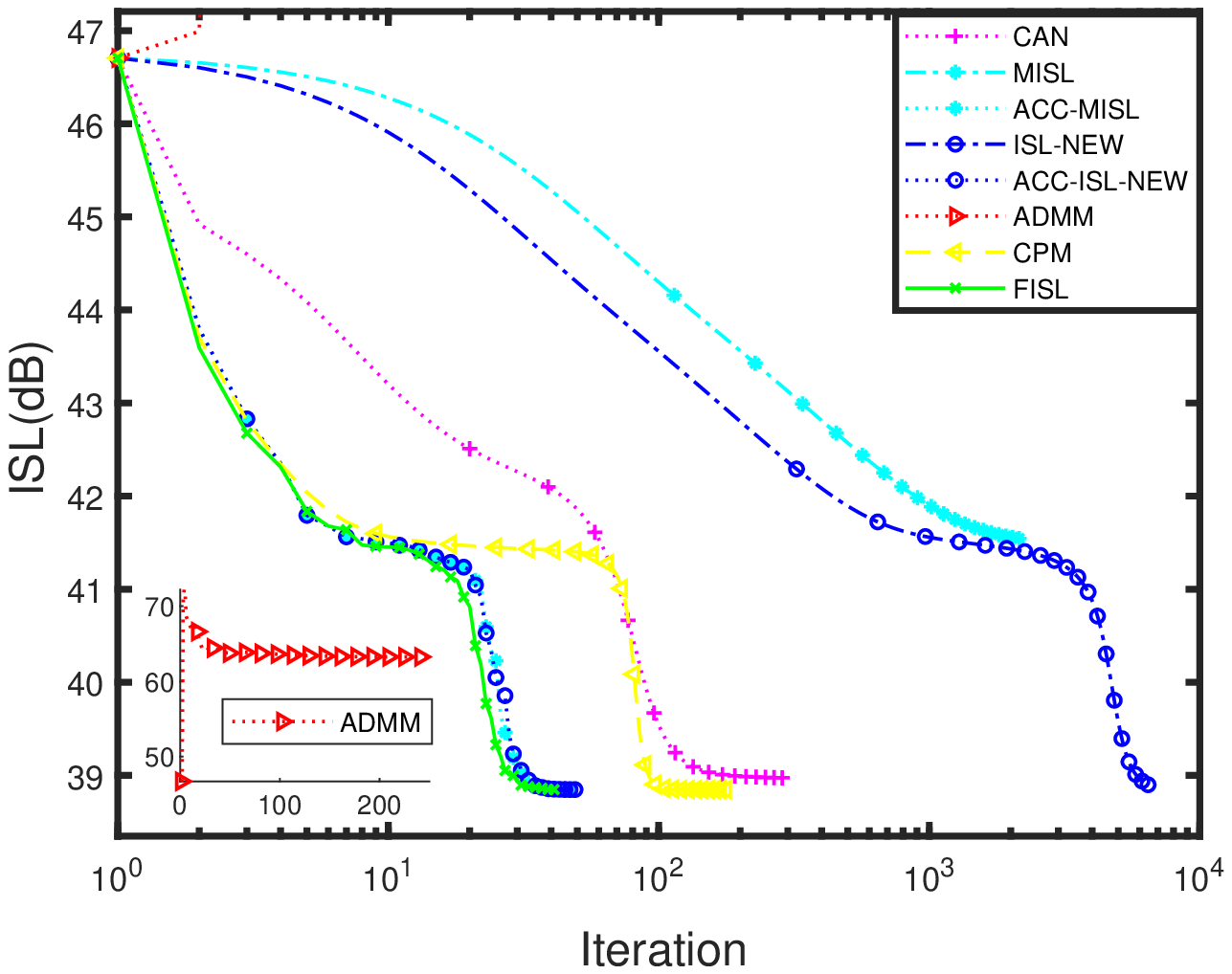}}\subfloat[$P=1225$]{\includegraphics[scale=0.55]{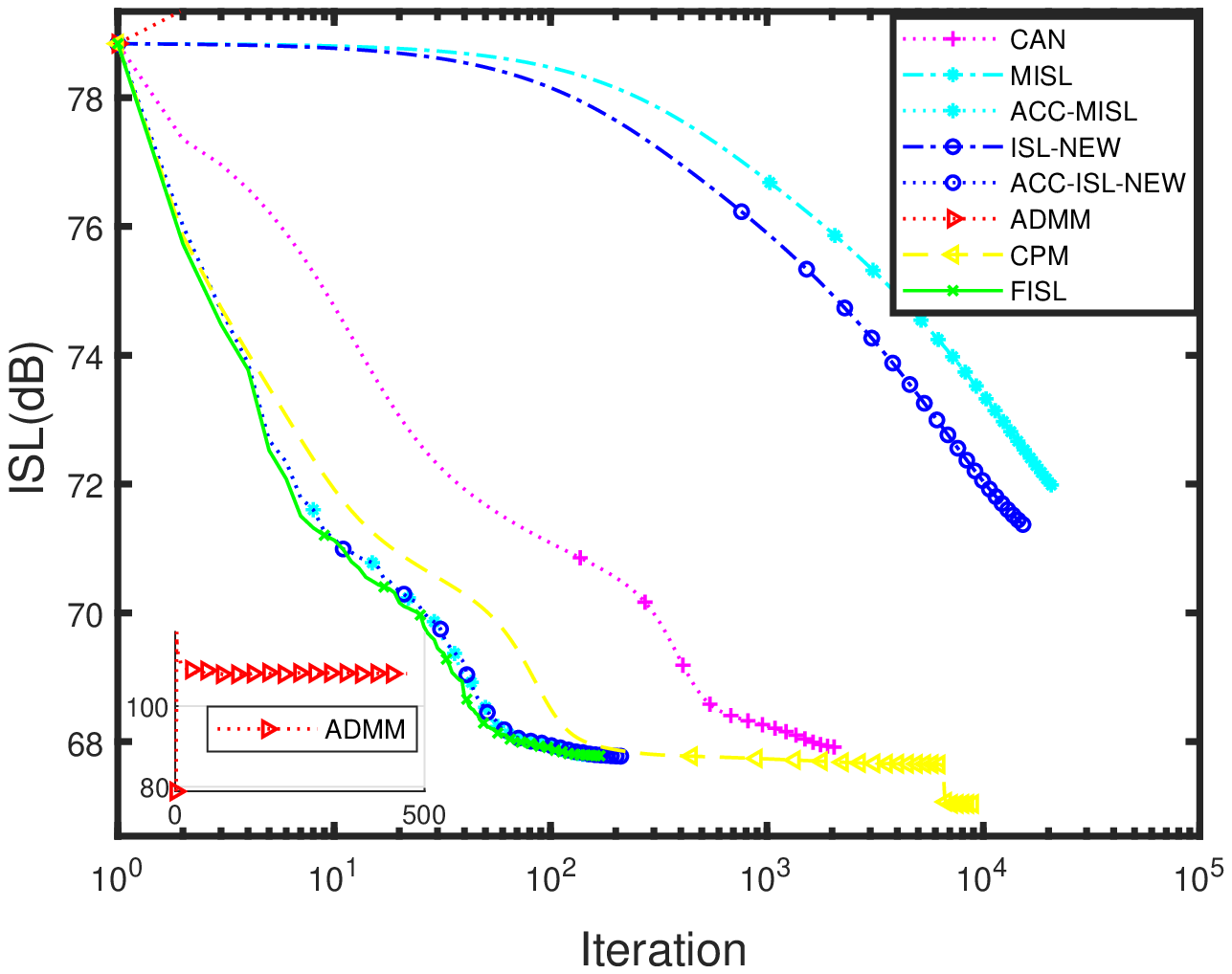}}

\caption{ISL with respect to iteration for a sequence length $P=100,1225$.
(a) and (b) are for initialization via Random sequence. (c) and (d)
are for initialization via Golomb sequence. (e) and (f) are for initialization
via Frank sequence.}
\end{figure}

\begin{figure}[tph]
\subfloat[$P=100$]{\includegraphics[scale=0.55]{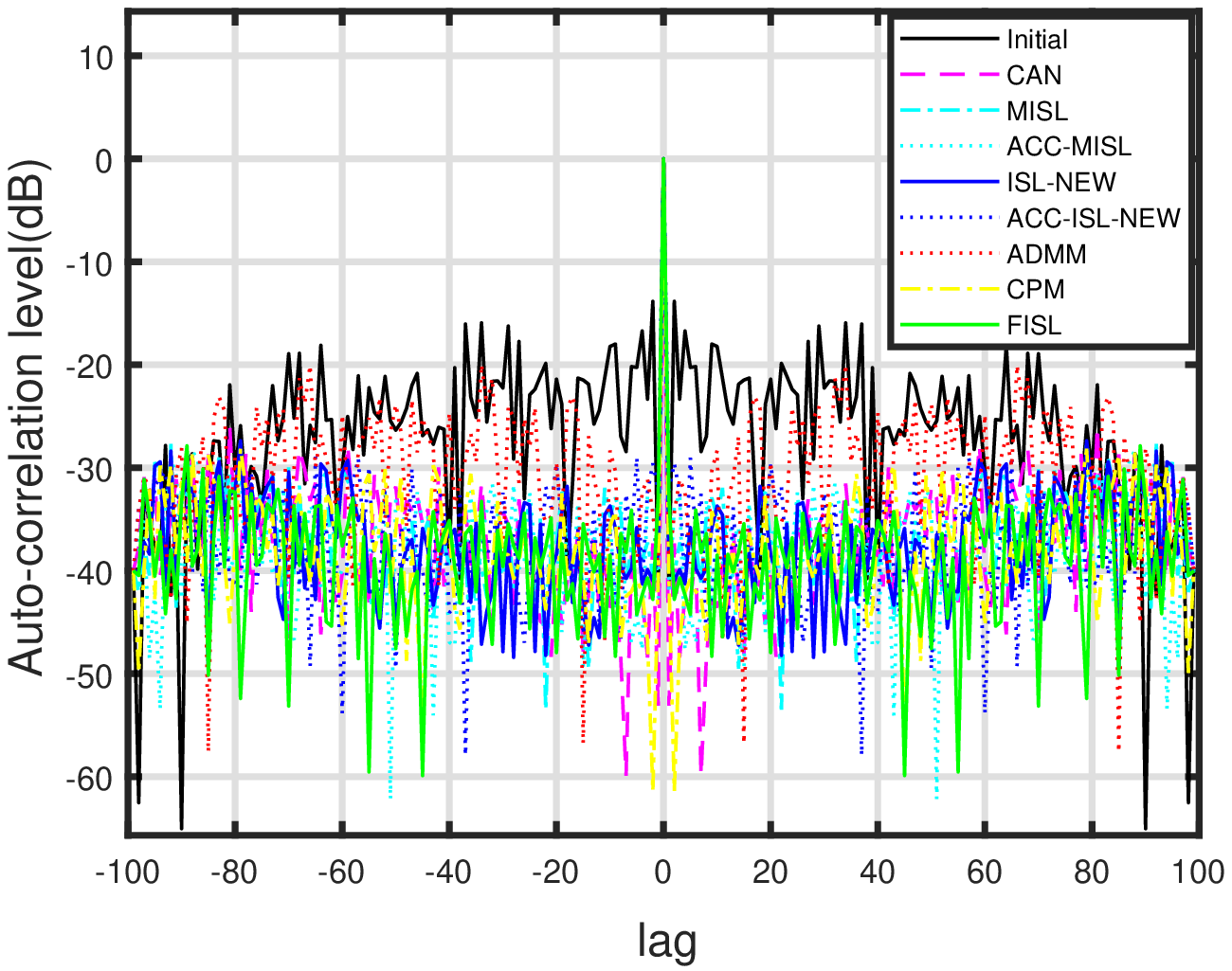}}\subfloat[$P=1225$]{\includegraphics[scale=0.55]{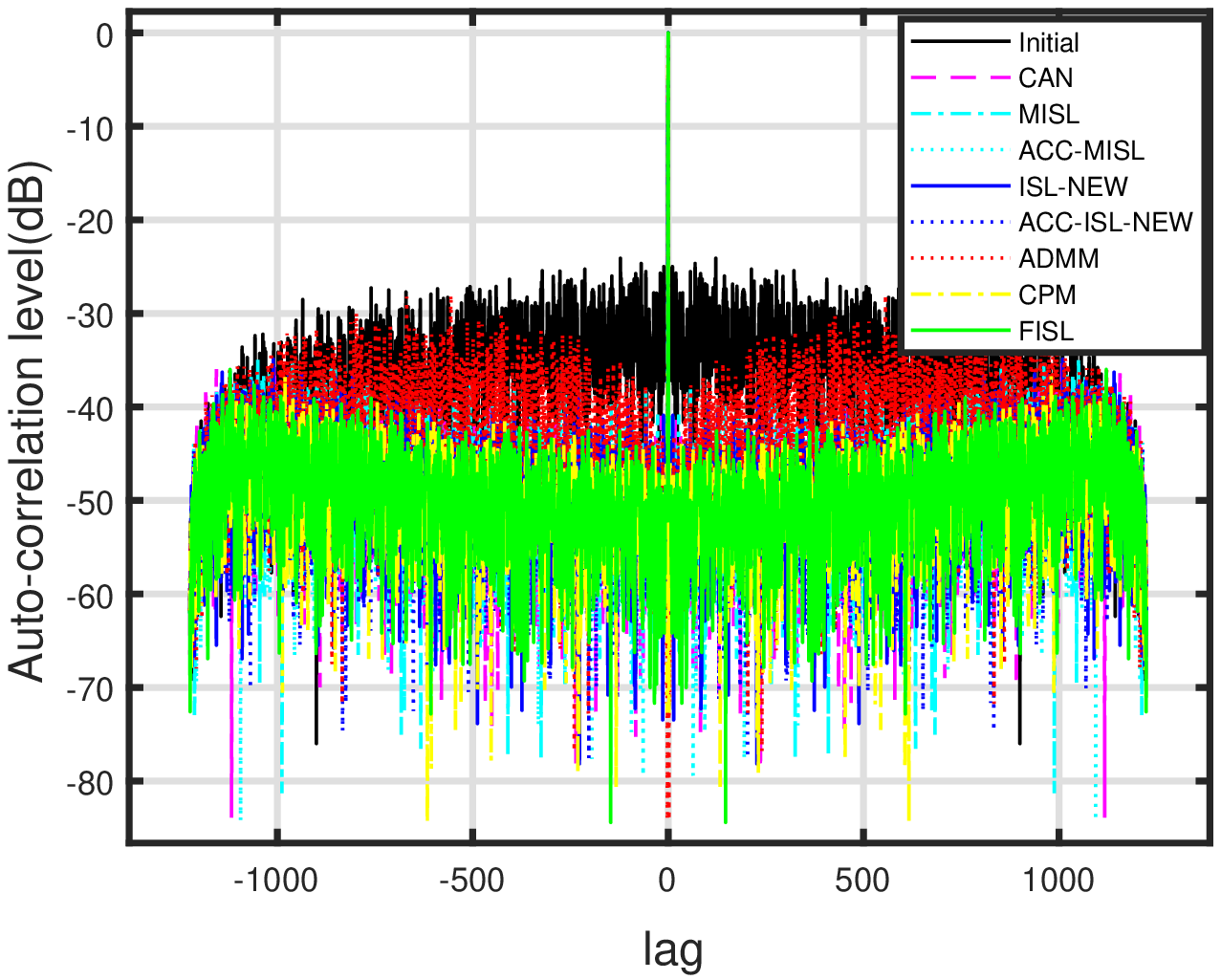}}

\subfloat[$P=100$]{\includegraphics[scale=0.55]{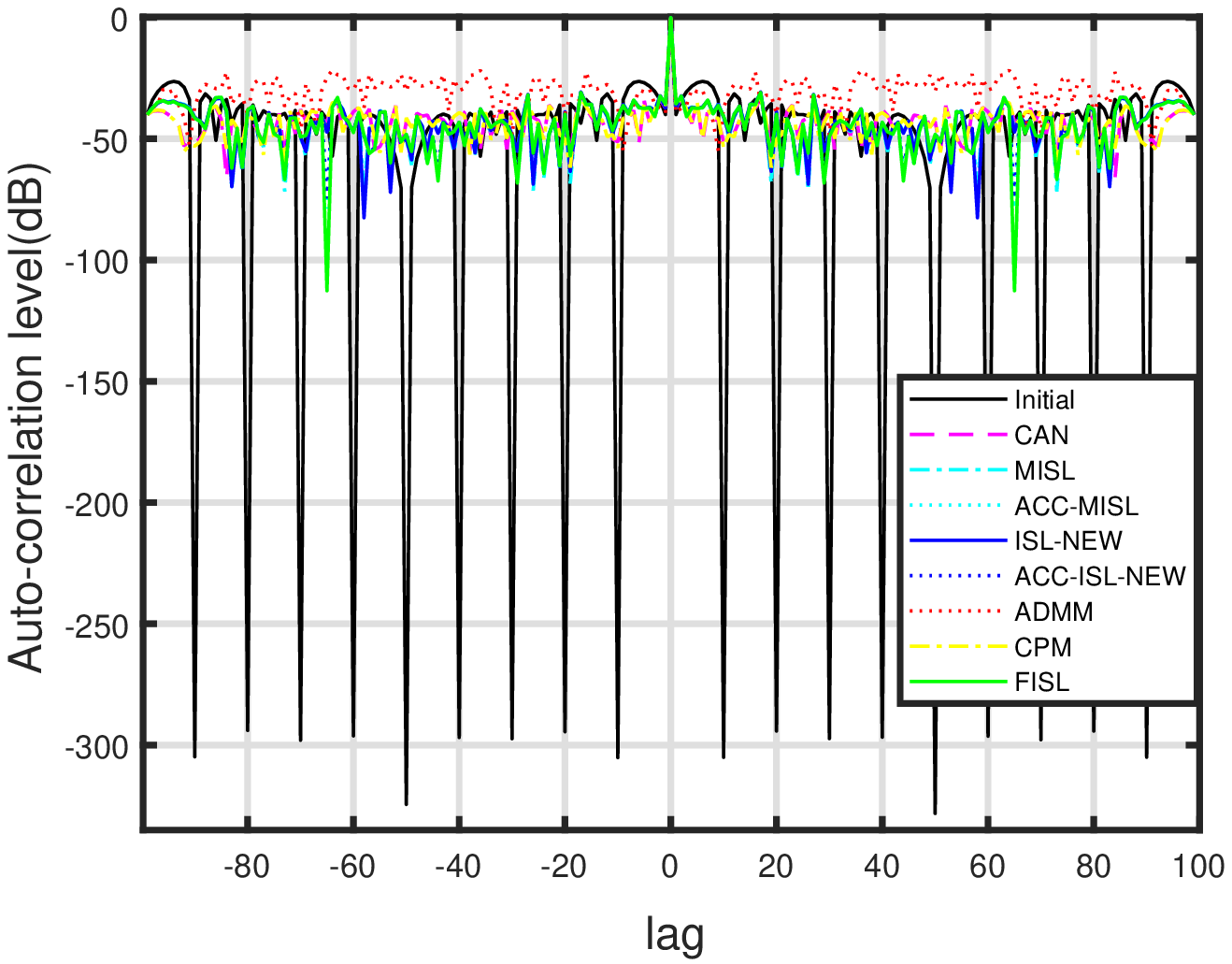}}\subfloat[$P=1225$]{\includegraphics[scale=0.55]{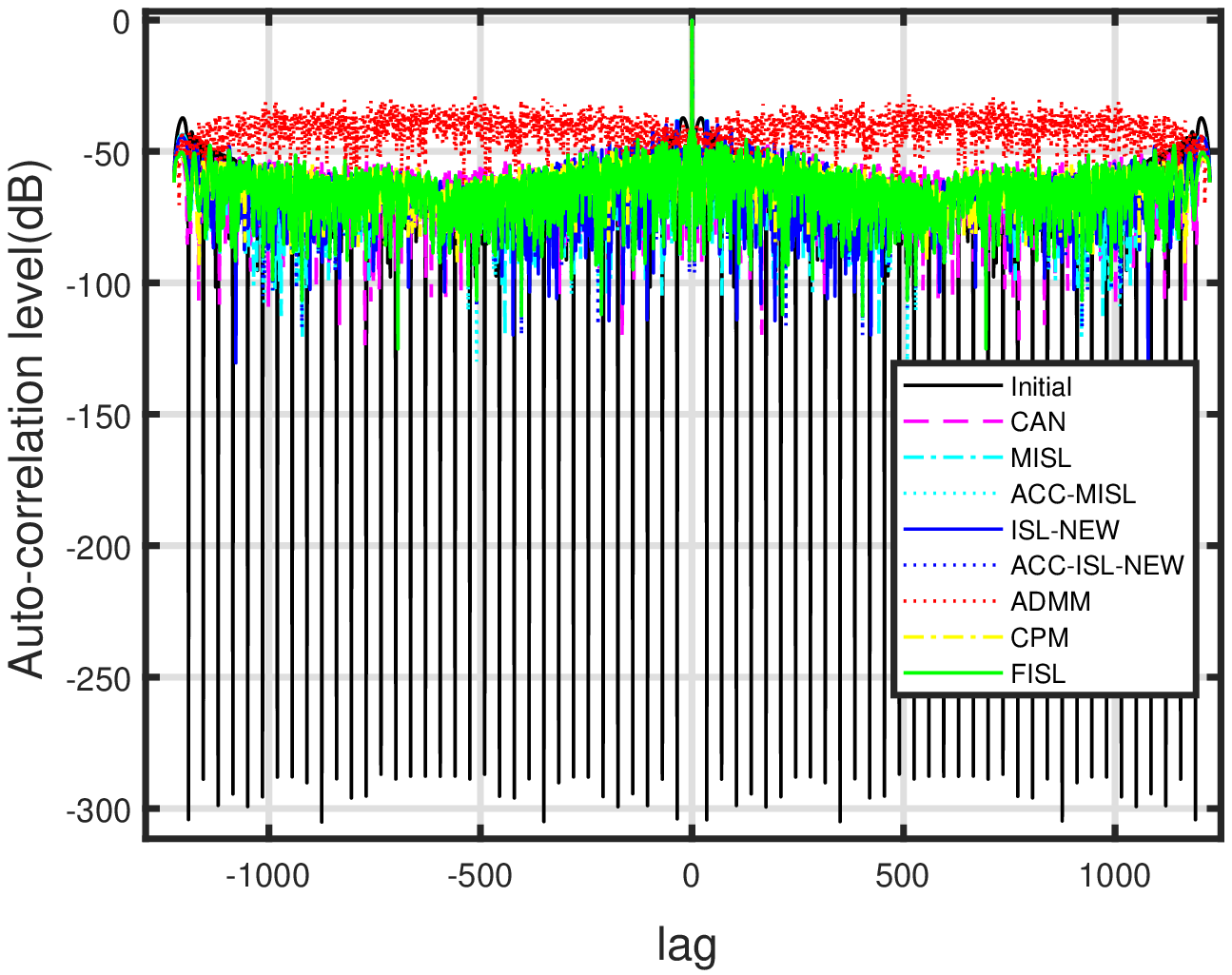}}

\subfloat[$P=100$]{\includegraphics[scale=0.55]{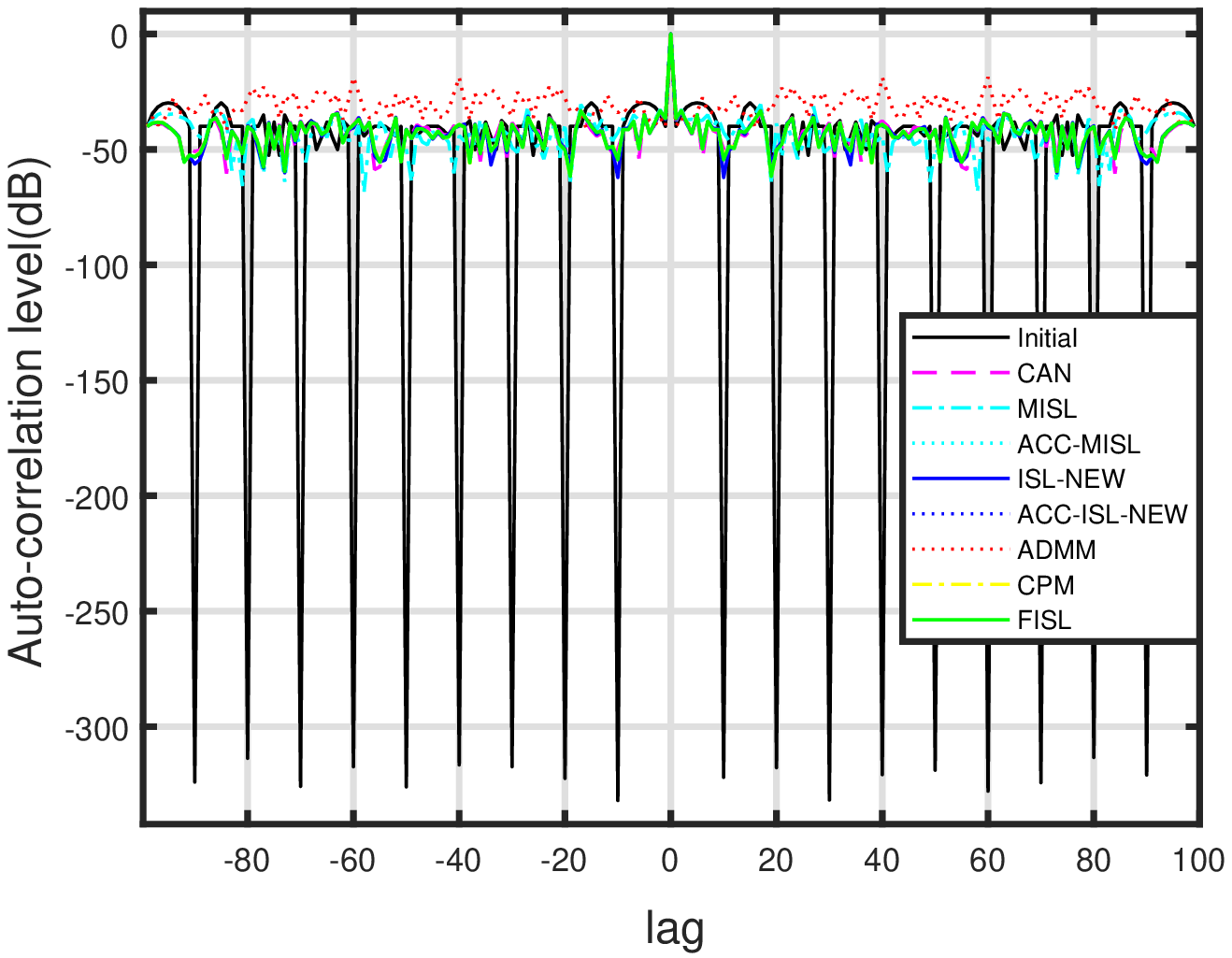}}\subfloat[$P=1225$]{\includegraphics[scale=0.55]{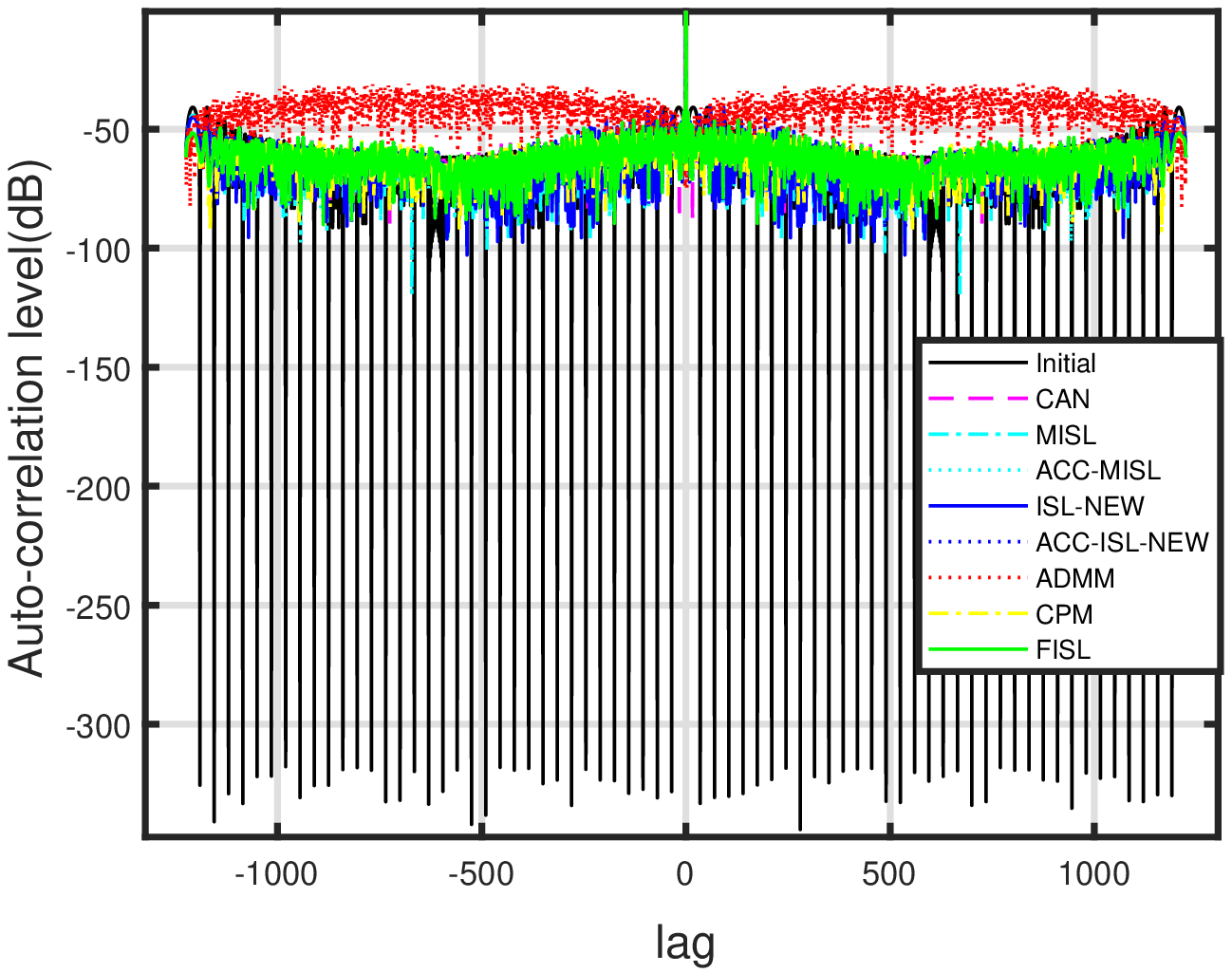}}

\caption{Auto-correlation value with respect to lag for a sequence length $P=100,1225$.
(a) and (b) are for initialization via Random sequence. (c) and (d)
are for initialization via Golomb sequence. (e) and (f) are for initialization
via Frank sequence.}
\end{figure}

\begin{figure}[tph]
\subfloat[for initialization via Random sequence]{\includegraphics[scale=0.55]{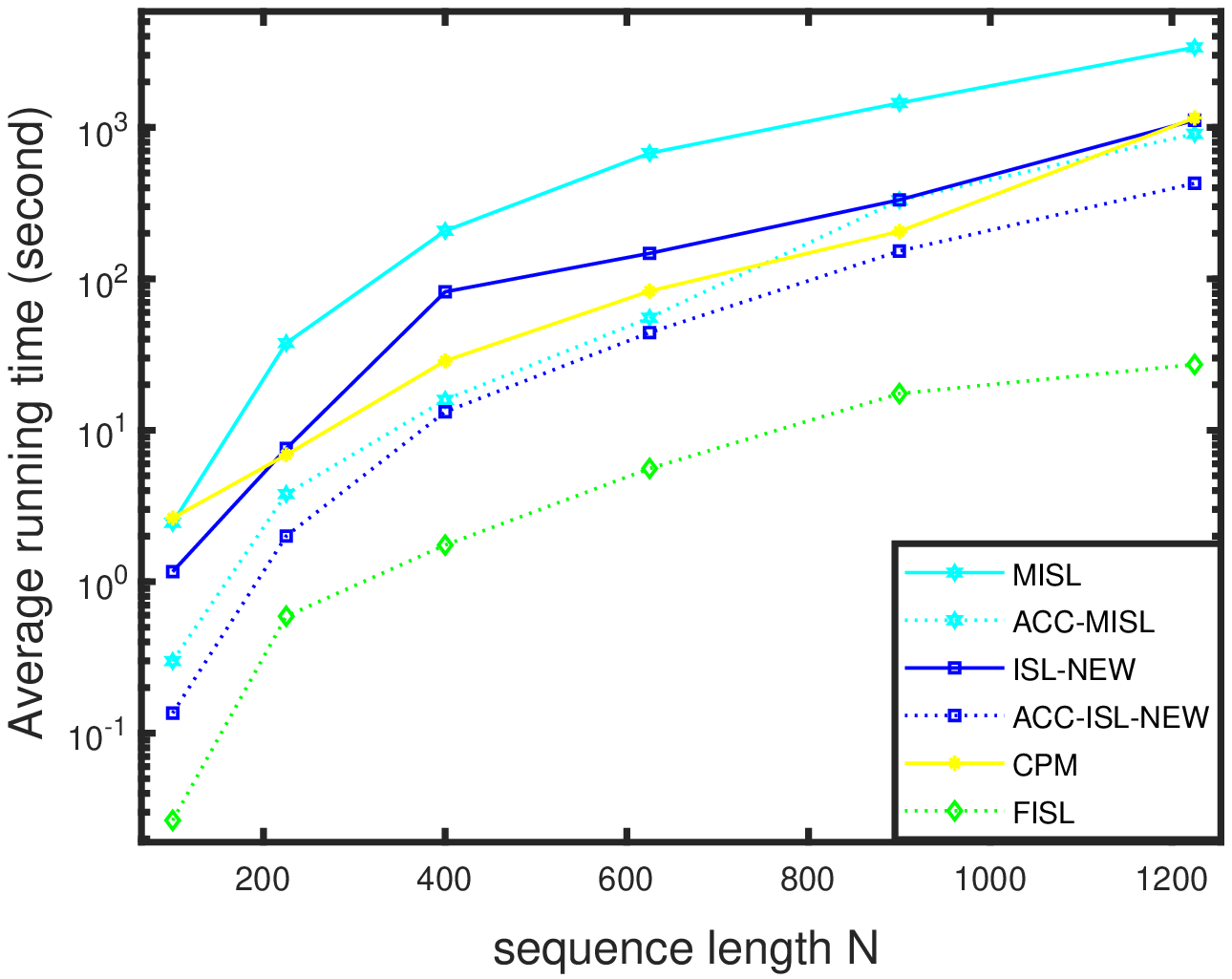}}\subfloat[for initialization via Golomb sequence]{\includegraphics[scale=0.55]{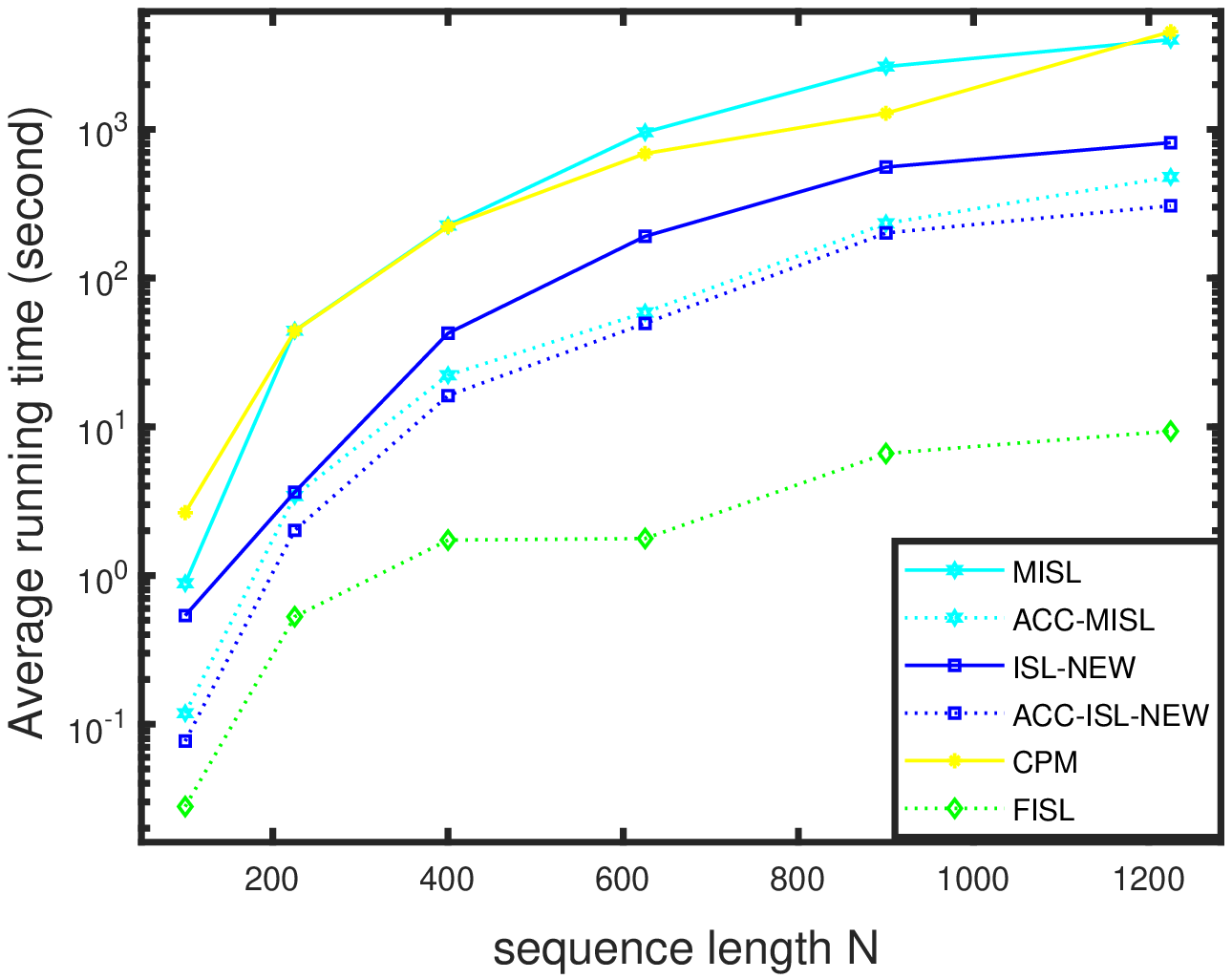}}

\caption{Average time with respect to sequence length}
\end{figure}

Figures. 3, 4 shows the normal and zoomed versions of the comparison
plots of ISL value vs time, ISL value vs the number of iterations
for different lengths and different initializations, respectively.
We have considered the squared iterative method (SQUAREM) \cite{17_MISL}
accelerating scheme to implement the accelerated MISL (ACC-MISL) and
accelerated ISL-NEW (ACC-ISL-NEW) algorithms. From simulation plots,
one can observe that all the algorithms are starting at the same objective
value, except the CAN and ADMM method all the methods are converging
to the same minimum value but with different converging rates. From
figures-3(b) and 4(b), for a sequence length of $P=1225$, FISL algorithm
is faster than the MISL, ACC-MISL, ISL-NEW, ACC-ISL-NEW and CPM algorithms
by $125,34,42,20,43$ times (with respect to the convergence time),
$123,14,119,10,9$ times (with respect to the number of iterations)
respectively.

Now in Figure. 5, we are comparing all the algorithms in terms of
auto-correlation side-lobe levels vs different lags, for different
sequence lengths and different initializations. From simulation plots,
we observe that except the ADMM approach, all the other algorithms
are performing well in terms of the PSL metric value.

Figure. 6 consists of the comparison plots of average running time
vs different sequence lengths for two different initializations. From
simulation plots, one can observe that, irrespective of the sequence
length and initialization, FISL algorithm is always taking less time
when compared to the state-of-the-art algorithms. From figure-6(a),
one can observe that the FISL algorithm is better than the MISL, ACC-MISL,
ISL-NEW, ACC-ISL-NEW, and CPM algorithms by $126,34,42,16,46$ times
respectively.

\section*{\centerline{IV.Conclusion}}

In this paper, we address the problem, design of phase only sequences
of arbitrary lengths by directly minimizing the ISL metric. We proposed
a fast iterative algorithm by using the Majorization-Minimization
method. Numerical simulations of the proposed algorithm were conducted
for different sequence lengths using different initializations that
confirm our algorithm performs better than the state-of-the-art algorithms
in terms of the speed of convergence.

\newpage{}

\bibliographystyle{IEEEtran}
\bibliography{FISL_Paper}

\end{document}